\documentclass[
  journal=pasa,
  manuscript=research-paper, 
  year=2020,
  volume=37,
]{cup-journal}
\newcommand{\gh}{GLEAM\,\-J0917\-$-$0012}
\usepackage{microtype,siunitx,booktabs}
\usepackage{hyperref}
\usepackage{amssymb}

\usepackage{acronym}

\acrodef{aips}[\texttt{AIPS}]{\texttt{Astronomical Image Processing System}}
\acrodef{agn}[AGN]{active galactic nucleus}
\acrodef{agns}[AGN]{active galactic nuclei}
\acrodef{alma}[ALMA]{Atacama Large Millimeter/submillimeter Array}
\acrodef{askap}[ASKAP]{Australian SKA Pathfinder Telescope}
\acrodef{atca}[ATCA]{Australia Telescope Compact Array}
\acrodef{askap}[ASKAP]{Australian SKA Pathfinder Telescope}
\acrodef{beagle}[\texttt{BEAGLE}]{\texttt{BayEsian Analysis of GaLaxy sEds}}
\acrodef{boss}[BOSS]{Baryonic Oscillation Spectroscopic Survey}
\acrodef{casa}[\texttt{CASA}]{\texttt{Common Astronomy Software Applications}}
\acrodef{cgm}[CGM]{circum-galactic medium}
\acrodef{cosmos}[COSMOS]{Cosmic Evolution Survey}
\acrodef{css}[CSS]{compact steep spectrum}
\acrodef{eelr}[EELR]{extended emission line region}
\acrodef{eelrs}[EELRs]{extended emission line regions}
\acrodef{efeds}[eFEDS]{eROSITA Final Equatorial Depth Survey}
\acrodef{emu}[EMU]{Evolutionary Map of the Universe}
\acrodef{erosita}[eROSITA]{extended ROentgen Survey with an Imaging Telescope Array}
\acrodef{fwhm}[FWHM]{full-width half maximum}
\acrodef{gama}[GAMA]{Galaxy and Mass Assembly}
\acrodef{GLEAM-X}[GLEAM-X]{GaLactic and Extragalactic All-sky MWA X}
\acrodef{gmrt}[GMRT]{Giant Metrewave Radio Telescope}
\acrodef{gps}[GPS]{giga-Hertz peaked spectrum}
\acrodef{grizli}[\texttt{grizli}]{\texttt{Grism Redshift and Line Analysis Software}}
\acrodef{hzrg}[HzRG]{high-redshift radio galaxy}
\acrodef{hzrgs}[HzRGs]{high-redshift radio galaxies}
\acrodef{hsc}[HSC]{Hyper Suprime-Cam}
\acrodef{hst}[\textit{HST}]{\textit{Hubble Space Telescope}}
\acrodef{icrf}[ICRF]{International Celestial Reference Frame}
\acrodef{ir}[IR]{infra-red}
\acrodef{ips}[IPS]{inter-planetary scintillation}
\acrodef{ism}[ISM]{interstellar medium}
\acrodef{lba}[LBA]{Long Baseline Array}
\acrodef{llagn}[LLAGN]{low-luminosity AGN}
\acrodef{lofar}[LOFAR]{LOw Frequency ARray}
\acrodef{mps}[MPS]{mega-Hertz peaked spectrum}
\acrodef{mwa}[MWA]{Murchison Widefield Array}
\acrodef{ne}[NE]{north-east}
\acrodef{nsi}[NSI]{normalised scintillation index}
\acrodef{nrao}[NRAO]{National Radio Astronomy Observatory}
\acrodef{nvss}[NVSS]{NRAO VLA Sky Survey}
\acrodef{pdf}[PDF]{probability distribution function}
\acrodef{pdfs}[PDFs]{probability distribution functions}
\acrodef{psf}[PSF]{point spread function}
\acrodef{qso}[QSO]{quasi-stellar object}
\acrodef{qsos}[QSOs]{quasi-stellar objects}
\acrodef{raise}[\texttt{RAiSE}]{\texttt{Radio AGN in Semi-analytic Environments}}
\acrodef{rms}[RMS]{Root Mean Squared error}
\acrodef{sdss}[SDSS]{Sloan Digital Sky Survey-III}
\acrodef{sed}[SED]{spectral energy distribution}
\acrodef{sf}[SF]{star formation}
\acrodef{sfr}[SFR]{star formation rate}
\acrodef{sfgs}[SFGs]{star forming galaxies}
\acrodef{smbh}[SMBH]{super-massive black hole}
\acrodef{smc}[SMC]{Small Magellenic Cloud}
\acrodef{snr}[SNR]{signal-to-noise ratio}
\acrodef{srg}[SRG]{Spektrum-Roentgen-Gamma}
\acrodef{summs}[SUMSS]{Sydney University Molonglo Sky Survey}
\acrodef{sw}[SW]{south-west}
\acrodef{tgss}[TGSS]{TIFR GMRT Sky Survey}
\acrodef{tifr}[TIFR]{Tata Institute of Fundamental Research}
\acrodef{uss}[USS]{ultra-steep spectrum}
\acrodef{uv}[UV]{ultra-violet}
\acrodef{vla}[VLA]{Very Large Array}
\acrodef{wfc3}[WFC3]{Widefield Camera 3}
\acrodef{wise}[\textit{WISE}]{\textit{Widefield Infrared Survey Explorer}}

\title{A Jet-Induced Shock in a Young, Powerful Radio Galaxy at $z=3.00$}

\author{N. Seymour}
\affiliation{International Centre for Radio Astronomy Research, Curtin University, Bentley, WA 6102, Australia}
\email[N. Seymour]{nick.seymour@curtin.edu.au}

\author{J.W. Broderick}
\affiliation{SKA Observatory, Science Operations Centre, CSIRO ARRC, 26 Dick Perry Avenue, Kensington, WA 6151, Australia}
\alsoaffiliation{CSIRO Space and Astronomy, PO Box 1130, Bentley, WA 6102, Australia}
\alsoaffiliation{International Centre for Radio Astronomy Research, Curtin University, Bentley, WA 6102, Australia}

\author{G. Noirot}
\affiliation{Institute for Computational Astrophysics and Department of Astronomy \& Physics, Saint Mary's University, 923 Robie Street, Halifax, NS B3H 3C3, Canada}
\alsoaffiliation{Space Telescope Science Institute, 3700 San Martin Drive, Baltimore, Maryland 21218, USA}

\author{R.J. Turner}
\affiliation{School of Natural Sciences, University of Tasmania, Private Bag 37, Hobart, 7001, Australia}

\author{A.J. Hedge}
\affiliation{International Centre for Radio Astronomy Research, Curtin University, Bentley, WA 6102, Australia}

\author{A. Gupta}
\affiliation{International Centre for Radio Astronomy Research, Curtin University, Bentley, WA 6102, Australia}

\author{C. Reynolds}
\affiliation{CSIRO Space and Astronomy, PO Box 1130, Bentley, WA 6102, Australia}

\author{Tao An}
\affiliation{Shanghai Astronomical Observatory, CAS, 80 Nandan Road, Shanghai 200030, China}
\affiliation{Key Laboratory of Radio Astronomy and Technology, Chinese Academy of Sciences, A20 Datun Road, Chaoyang District, Beijing 100101, China}

\author{B. Emonts}
\affiliation{ National Radio Astronomy Observatory, 520 Edgemont Road, Charlottesville, VA 22903, USA}

\author{K. Ross}
\affiliation{International Centre for Radio Astronomy Research, Curtin University, Bentley, WA 6102, Australia}

\author{D. Stern}
\affiliation{Jet Propulsion Laboratory, California Institute of Technology, 4800 Oak Grove Drive, Mail Stop 264-789, Pasadena, CA 91109, USA}

\author{J. M. Afonso}
\affiliation{Instituto de Astrof\'{i}sica e Ci\^{e}ncias do Espa\c co, Universidade de Lisboa, OAL, Tapada da Ajuda, PT1349-018 Lisboa, Portugal}
\alsoaffiliation{Departamento de F\'{i}sica, Faculdade de Ci\^{e}ncias, Universidade de Lisboa, Edif\'{i}cio C8, Campo Grande, PT1749-016 Lisbon, Portugal}

\doi{10.1017/pasa.2020.32}

\received {dd Mmm YYYY}
\revised  {dd Mmm YYYY}
\accepted {dd Mmm YYYY}
\published{22 September 2020}


\begin{document}

\begin{abstract}
The bright radio source, GLEAM J091734$-$001243 (hereafter \gh), was previously selected as a candidate ultra-high redshift ($z>5$) radio galaxy due to its compact radio size and faint magnitude ($K(\rm AB)=22.7$). Its redshift was not conclusively determined from follow-up millimetre and near-infrared spectroscopy. Here we present new {\it HST} WFC3 G141 grism observations which reveal several emission lines including [NeIII]$\lambda$3867, [NeV]$\lambda$3426 and an extended ($\approx$$4.8\,$kpc), [OII]$\lambda$3727 line which confirm a redshift of $3.004\pm0.001$. The extended component of the [OII]$\lambda$3727 line is co-spatial with one of two components seen at 2.276\,GHz in high resolution ($60\times 20\,$mas) Long Baseline Array data, reminiscent of the alignments seen in local compact radio galaxies. The {\tt BEAGLE} stellar mass ($\approx$$2\times 10^{11}\,$M$_\odot$) and radio luminosity ($L_{\rm 500MHz}\approx$$10^{28}\,$W\,Hz$^{-1}$) put \gh~within the distribution of the brightest high-redshift radio galaxies at similar redshifts. However, it is more compact than all of them. Modelling of the radio jet demonstrates that this is a young, $\approx 50\,$kyr old, but powerful, $\approx 10^{39}\,$W, compact steep spectrum radio source. The weak constraint on the active galactic nucleus bolometric luminosity from the [NeV]$\lambda$3426 line combined with the modelled jet power tentatively implies a large black hole mass, $\ge 10^9\,$M$_\odot$, and a low, advection-dominated accretion rate, {\bf i.e. an} Eddington ratio $\le 0.03$. The [NeV]$\lambda$3426/[NeIII]$\lambda$3867 vs [OII]$\lambda$3727/[NeIII]$\lambda$3867 line ratios are most easily explained by radiative shock models with precursor photoionisation. Hence, we infer that the line emission is directly caused by the shocks from the jet and that this radio source is one of the youngest and most powerful known at cosmic noon. We speculate that the star-formation in \gh~could be on its way to becoming quenched by the jet.
\end{abstract}

\section{INTRODUCTION }
\label{sec:int}

Over the last few decades `feedback' processes between the central black hole and host galaxy have become part of the accepted paradigm for galaxy evolution. Numerical simulations were initially found to over-produce the number of massive galaxies \citep{Bower:06,Croton:06} until a mechanism was introduced to cool gas and prevent excessive star formation at the high mass end of the galaxy stellar mass function. This mechanism was proposed to be energy injected into the \ac{cgm} and intracluster medium by jets from the central black hole (although feedback may occur through means other than radio jets). Now there are numerous observations which support the idea that radio jets produced by \ac{agn} provide one of the main forms of mechanical feedback. This negative `feedback' occurs on both large scales \citep[e.g. radio jets creating cavitites in cluster X-ray halos;][]{Gitti:11} and small scales \citep[e.g. as demonstrated by the alignment of extended emission lines with jets in \ac{css} radio sources,][]{devries:97,Axon:08}. Indeed, the compact radio-loud QSO 3C48 at $z=0.369$ is the archetypal \ac{css} source presenting alignment between the optical and compact ($<5\,$kpc) radio emission \citep{Stockton:07,An:10}.

The population of compact radio galaxies known as \ac{gps} and \ac{css} sources \citep{Peacock:82} are prime candidates for jet feedback within the \ac{ism} of the host galaxy \citep{odea:21}. \ac{gps} sources are characterised as having  radio spectra peaking between 1-5\,GHz in the observer's frame \citep{Gopal-Krishna:83} and tend to have sizes $<500\,$pc, but the broader class of peaked radio sources can have a peak outside typical observing frequencies depending on their redshift and age. \ac{css} sources are compact radio sources that have steep spectra ($\alpha<-0.5$, where the flux density $S_\nu\propto\nu^\alpha$ and $\nu=$ frequency) over the entire observed frequency range, and are observed to be slightly larger than \ac{gps} sources, $0.5-20\,$kpc. As the rest-frame peak frequency of \ac{gps} sources decreases with increasing source size \citep{odea:97} the natural inference is that \ac{css} sources are older or higher-redshift sources where the peak frequency has been shifted below the range of current observations. Indeed, radio sources with measurable peaks below $400\,$MHz have been referred to as \ac{mps} sources \citep[e.g.][]{Callingham:17,Ross:22} and show evidence for a redshift distribution up to, and possibly beyond, $z\approx 2.4$ \citep{Coppejans:15}.

Evidence that \ac{gps}/\ac{css} sources are involved in feedback including (a) the kinematics of the jet-aligned extended emission lines, (b) the asymmetry of radio morphology suggesting interaction with dense clouds in the \ac{ism} \citep[e.g.][]{Saika:95} and (c) observations of hot gas from X-ray observations which is also shocked and aligned with the jet \citep[e.g.~][]{Massaro:09}. A common feature of extended, powerful radio galaxies at low-redshift is the presence of \ac{eelrs} on scales of $11-45\,$kpc \citep{Fu:09a}. While they are often seen with velocities of $500-1100\,$km\,s$^{-1}$ with indications that they are photoionised by the \ac{agn}, they do not show any alignment with the radio jet \citep{Fu:09b}. This result is in contrast to \ac{css} sources which have direct evidence that they are caused by shocks \citep{Fu:06}. The inference is that there is evolution from aligned \ac{eelrs} in more compact ($\lesssim 10\,$kpc) radio sources due to shocks to a more disorderly alignment in more extended radio sources due to photoionisation \citep{Shih:13}. 

Studying radio galaxies at higher redshift provides a window into many galaxy evolution processes, e.g. the co-evolution of the galaxies with their central black hole \citep[e.g.][]{Drouart:16}, their proto-cluster environments \citep[e.g.][]{Wylezalek:13} and the interaction of the jets with the dense \ac{ism}, improving our knowledge of physics under extreme conditions \citep[e.g.][]{Ighina:22}. In fact, high-redshift radio galaxies often show intriguing alignments between the radio source and large-scale gaseous and stellar structures in the host galaxy and surrounding environment, such as those seen in rest-frame \ac{uv} light \citep{Pentericci:99}, submillimeter emission \citep{Stevens:03}, and cold molecular gas \citep{klamer:04,emonts:23}. While a possible explanation is that these alignments are caused by positive \ac{agn} feedback in the form of gas cooling and triggered star formation \citep[e.g.,][]{Gullberg:16,Nesvadba:20}, kinematically perturbed and shocked gas is often seen in the ionized gas along the extent of the radio jet \citep[e.g.,][]{VillarMartin:03}, and negative feedback has been implied from observations of jet-induced outflows \citep[e.g.,][]{Nesvadba:17}.

Understanding \ac{agn} feedback mechanisms in these epochs is critical to unravelling how these processes shaped galaxy evolution. From a sample of 3CR galaxies at $z\approx 1$, \cite{Best:00b} demonstrated that smaller radio galaxies ($\leq 120\,$kpc) were more likely to have larger \ac{eelrs} which are powered by shocks and have have kinematic velocities $>1000\,$km\,s$^{-1}$. The larger radio galaxies in this sample have \ac{eelrs} powered by photoionisation instead. The same trend of high velocity, shock ionisation of \ac{eelrs} in compact, $< 120\,$kpc, radio galaxies was also seen by \cite{DeBreuck:00} in a sample of powerful radio galaxies across $0<z<5.2$. \ac{gps} and \ac{css} sources are therefore the most extreme of these powerful radio sources consistent with the idea that the photoioinisation of their \ac{eelrs} is caused by jet-induced shocks.

A new generation of low frequency radio surveys such as the \ac{tifr} \acl{gmrt} \citep[\acs{gmrt};][]{Swarup:91} Sky Survey (\acs{tgss}) Alternative Data Release 1 \citep[][]{Intema:17}, the \acl{lofar} \citep[\acs{lofar};][]{VanHaarlem:13} Two-metre Sky Survey \citep[LoTSS;][]{Shimwell:17} and the GaLactic and Extragalactic All-sky \acl{mwa} \citep[\acs{mwa};][]{Tingay:13} survey \citep[GLEAM;][]{Wayth:15} has inspired several new searches for high redshift radio galaxies. While several searches rely on the classical \acl{uss} (\acs{uss}, $\alpha<-1.3$) technique \citep[e.g.][]{Saxena:18b,Gloudemans:22} the broad 72--231\,MHz frequency coverage of the GLEAM survey allows the shape of the radio spectra to be used as a selection tool. \citet[][herefter D20]{Drouart:20} used spectral curvature in the GLEAM band, non-detections in the VISTA Kilo-degree Infrared Galaxy (VIKING) survey \citep{Edge:13}, and compact radio sizes at high-frequency as a selection method finding one out of four bright, $S_{\rm 151MHz}>100\,$mJy, radio galaxies lying at $z=5.5$. 

The motivation for choosing sources with curvature at low frequency comes from empirical and theoretical modelling of the radio spectra of powerful radio galaxies at $z\approx 2-4$ \citep[D20][]{Broderick:22}. However, one can see how this method would select sources with peaked spectra within or below the GLEAM band (i.e. high redshift \ac{mps} or \ac{css} sources) and follows the hypothesis that $z>7$ radio galaxies will present redshifted peaked spectra due to the denser environment within the \ac{ism}/\ac{cgm}, more intense radiation from the cosmic microwave background as well as their age. From the handful of \ac{qsos} detected in the radio at $z>6$, several show evidence for flattening of their radio spectra at lower frequencies \citep{Banados:21,Ighina:22,Endsley:22}. D20 therefore confirmed that selecting sources with extreme radio to $K$-band flux density ratios, $S_{\rm 151MHz}/S_{2\mu{\rm m}}>10^5$, and faint $K$-band magnitudes is an effective selection technique to find $z>5$ radio galaxies \citep[e.g.][]{Broderick:24}.

Another source from D20, GLEAM J091734$-$001243 (hereafter \gh), was also thought to potentially be at high redshift.  As well as being compact, \gh~has a faint $K$-band counterpart - both characteristics of very distant radio galaxies. However, follow-up millimetre \citep[][hereafter D21]{Drouart:21} and near-infrared spectroscopy \citep[][hereafter S22]{Seymour:22}, failed to conclusively determine its redshift, necessitating the current study. Modelling in D21 demonstrated that the ALMA 100\,GHz continuum was synchrotron dominated. In S22, {\ac{hst}} \ac{wfc3} grism observations detected one weak line at $1.15\,\mu$m and  continuum detections of the host galaxy in the  F098M and F105W bands. A nearby, $\approx 1.5''$ offset,  source was also identified (see Fig.~\ref{fig:overlay}). Broad-band \ac{uv} to radio \ac{sed} fitting in S22 constrained by the $K$-band, the two \ac{wfc3} detections and the 100\,GHz photometry found two photometric redshift solutions at $z\approx 3$ and $z\approx 8$ consistent with the faint $1.15\,\mu$m line being MgII or Lyman-$\alpha$, respectively. 

In this paper, we present new \ac{wfc3} and southern hemisphere \ac{lba} data which demonstrate that \gh~is an extreme analogue of low redshift \ac{css} sources with strong evidence of a jet-induced shock. Our new observations are presented in \S\ref{sec:obs} with the results shown in \S\ref{sec:res}. We discuss these results in \S\ref{sec:dis} before presenting our conclusion in \S\ref{sec:con}. Throughout this paper, we quote magnitudes in the AB system \citep{Oke:83}, adopt a flat $\Lambda$CDM cosmology with parameters $h = 0.7$, $\Omega_{\rm M} = 0.3$, and $\Omega_\Lambda = 0.7$, and report $68\%$ confidence interval (or $1\sigma$) uncertainties unless otherwise specified.

\section{OBSERVATIONS AND DATA PROCESSING}
\label{sec:obs}

\subsection{New \ac{hst} \ac{wfc3} Observations}
\label{subsec:hst}

We observed \gh~with \ac{wfc3} using the G141 grism and the F140W band for pre-imaging under proposal ID 16662. Our four orbits were observed in March and April 2022 and each orbit comprised four pairs of pre-imaging plus grism exposures. The orbits were paired into two different orientations chosen to avoid contamination of the host galaxy spectrum by the nearby source. These data complemented those presented in S22 (ID 16184) which comprised one orbit of F105W imaging and four orbits of G102 grism plus F098M pre-imaging (also paired into two different orientations). We processed the data using the \ac{grizli} 1.4.0.dev31 \citep{Brammer:19}. We reprocessed both the S22 {\bf observations} (F098M, F105W, G102) and new F140W+G141 ones following the same procedure as described in S22. We provide a brief description below.

The initial steps include direct imaging and grism exposure associations, flat-fielding of the exposures, pixel flagging (bad pixels and cosmic-rays rejection and persistence masking), relative exposure-level astrometric alignment, sky-background subtraction, and drizzling of the direct imaging (with $0.06\times 0.06\,$ arcsec pixels) and grism visits. After these initial steps, individual exposures are aligned to the {\it Gaia} DR3 catalogue \citep{Gaia:23}, a direct F098M+F105W+F140W image mosaic is created and diffraction spikes of bright sources are masked. Source detection is then performed on the mosaic image with the {\tt Source Extractor} \citep{Bertin:02} python wrapper {\tt sep} \citep{Barbary:16}, and matched-aperture photometry is performed on the three filters. Spectral traces are then identified based on the source catalogue and the spectral continua of the sources are modelled for contamination removal. After these steps, we then extract the 2D cutout of the spectra of the host and the nearby source and then use \ac{grizli} to fit the 2D G102+G141 grism data. 

\begin{figure*}[t]
\centering
\includegraphics[trim={1.2cm 5.1cm 0.0cm 3.9cm},clip,width=9.0cm]{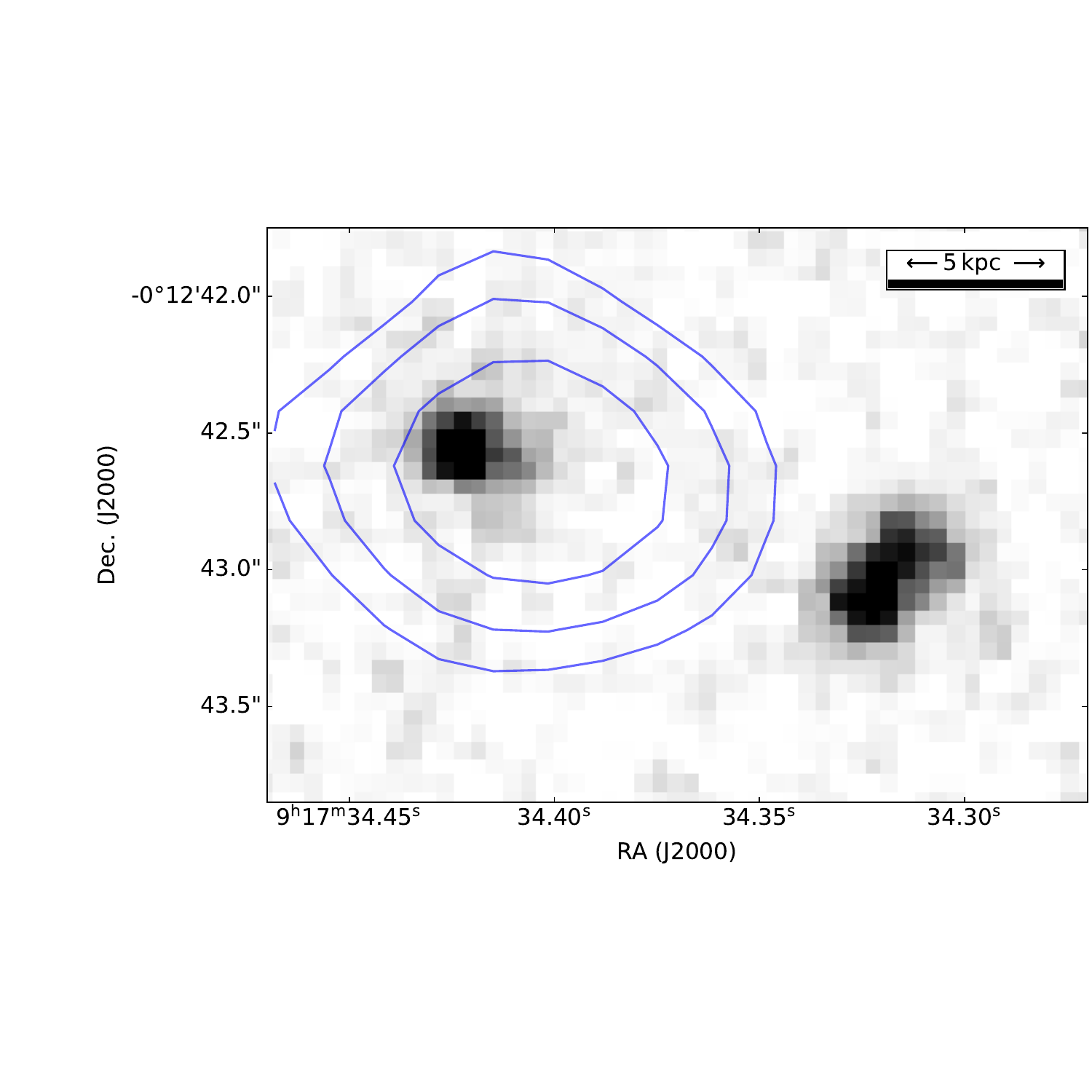}
\includegraphics[trim={0.2cm 4.1cm 0.5cm 3.9cm},clip,width=9.0cm]{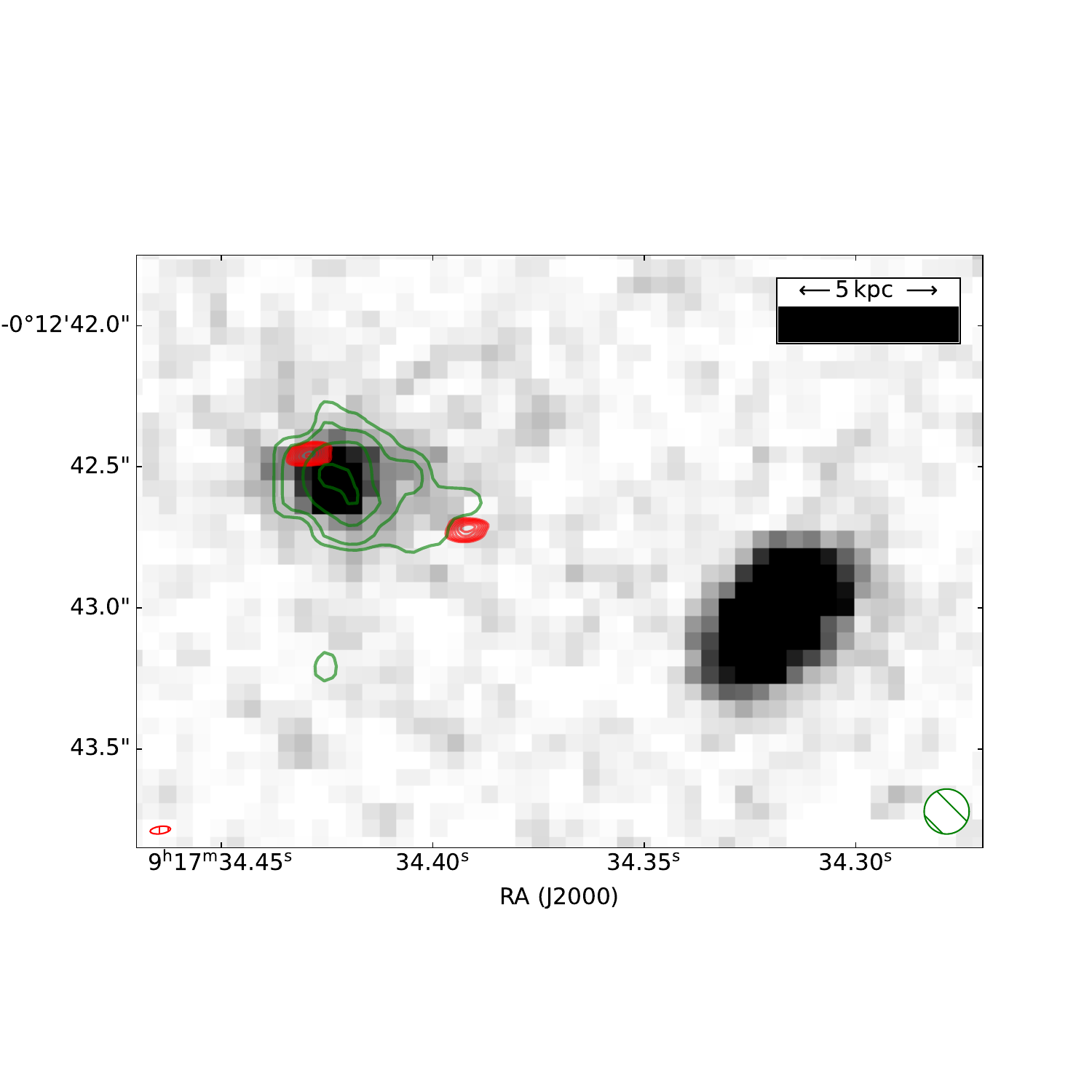}
\caption{
({\it left}) Inverted greyscale F140W image overlaid with the \ac{alma} 100\,GHz continuum (blue contours starting at $2.5\sigma$ with increasing steps of $\sqrt{2}$). The \ac{alma} data is unresolved with a beam of $1.2''\times 1.4''$ and position angle of $-89^\circ$ and is offset from the host galaxy. A bar in the top right of each images indicates the physical scale at $z=3.004$.
({\it right}) 
Inverted greyscale combined F140W/F105W/F098M image overlaid with the resolved [OII]$\lambda$3727 emission of the host (green contours, \ac{fwhm} of \ac{psf} in lower right) and the $2.276\,$GHz LBA data (red contours with the restoring beam in the lower left). The [OII]$\lambda$3727 contours start from $2.8\sigma$ and the \ac{lba} contours from $6\sigma$, both with increasing steps of $\sqrt{2}$. The extended [OII]$\lambda$3727 and the two components of the \ac{lba} data are aligned with the extension seen in the F140W band in the left panel although the alignment with the unrelated nearby source to the southwest is serendipitous.}
\label{fig:overlay}
\end{figure*}

In Table~\ref{tab:pos} we present the positions of the host and nearby source from \ac{grizli}. These values are the same as reported in S22. The optical photometry for the host is presented in Table~\ref{tab:hst}. We use the {\tt photutils} of {\tt astropy} to determine the photometry with a $0.7\,$arcsec radius aperture and a local background determined from a $3-5\,$arcsec annuli centred on the \ac{grizli} reported position.  Following S22, we determined the aperture corrections for \ac{wfc3}, by taking the inverse of the nearest set of values of the fractional flux enclosed in Table 7.6 of the \ac{wfc3} instrument handbook and interpolated them linearly using the $0.7\,$arcsec radius and central wavelength of each filter. A $5\%$ systematic uncertainty was added to the measurement uncertainty (to account for the absolute flux uncertainty and any uncertainty in the aperture correction). In Fig.~\ref{fig:overlay} we present the near-IR imaging along with the radio data described below. The spectroscopic results, including the 1D and 2D G141 spectra plus the \ac{grizli} best fits, are shown in Fig.~\ref{fig:spec:host} for the host and 
Fig.~\ref{fig:spec:com} for the nearby source. Note the absolute positional uncertainty of an \ac{hst} single pointings is $2-5\,$mas\footnote{\url{https://hst-docs.stsci.edu/drizzpac}}.

\begin{table}[t]
\begin{threeparttable}
\caption{Positions of the host of \gh~(Host) and the nearby galaxy as measured by \ac{grizli}.}
\label{tab:pos}
\begin{tabular}{@{}ccc@{}} \toprule
galaxy & R.A.(J2000) & dec.(J2000) \\
\midrule
host             &  09h17m34.42s & -00d12m42.5 \\
nearby galaxy    & 09h17m34.32s &  -00d12m43.0s \\
\bottomrule
\end{tabular}
\end{threeparttable}
\end{table}

\subsection{\ac{lba} Observations}
\label{subsec:lba}

We observed \gh~with the Southern Hemisphere \ac{lba} under project V605a from 2021 July 17 23:00 to 2021 July 18 09:00 UTC. The following telescopes were available: Murriyang (the 64-metre Parkes radio telescope), the \acl{atca} (\acs{atca}; five 22-m antennas in tied-array mode), Mopra, Hobart (26\,m), Ceduna, Warkworth (12\,m), Katherine (flagged due to poor sensitivity), Yarragadee, Hartebeesthoek (26\,m) and Tidbinbilla (70\,m). Target scans of duration 4\,min were interleaved with 1.5-min scans of the phase calibrator B0922$+$005. We also regularly observed B0906$+$015 to phase up \acs{atca}, used the sources B0208$-$512, B0537$-$441, B0834$-$201 and J1147$-$3812 for fringe finding, and observed B1934$-$638 for \acs{atca} flux density calibration. The observations were dual-polarisation, with four sub-bands centred at 2252, 2268, 2284 and 2300\,MHz, each with bandwidth 16\,MHz (i.e. a centre frequency of 2276\,MHz and a total bandwidth 64\,MHz). The total time on-source for \gh~was 5.6\,h.      

To correlate the data we made use of the DiFX software correlator \citep[][]{deller11} running at the Pawsey Supercomputing Centre. The data were averaged to 2\,s accumulation periods and 0.5\,MHz resolution during correlation. Parkes recorded native linear polarisations which were converted to a circular basis after correlation using the PolConvert package \citep[][]{2016A&A...587A.143M}. Data were calibrated in the \ac{aips}\footnote{\url{www.aips.nrao.edu}} following standard procedures for the \ac{lba}, implemented using the Parseltongue interface \citep[][]{2006ASPC..351..497K}. In the absence of good system temperature measurements at several stations, array amplitude calibration was achieved by bootstrapping to the compact source B0922$-$005 whose flux on \ac{lba} scales was assumed to be 90\% of that measured in a near-simultaneous \acs{atca} observation -- this assumption leads to a $\approx 10$\% uncertainty on the flux scale for the target source (this error estimate is supported by flux densities derived for calibrators in the observation that also had near-simultaneous \acs{atca} observations). Katherine had significantly reduced sensitivity due to a problem with the data acquisition system and was thus omitted from the final analysis.

The source was imaged using the \acl{casa} \citep[\acs{casa};][]{casa:22} with Briggs weighting  of $R=0.5$, a uvtaper of $5000\,$k$\lambda$ and the multi-frequency synthesis option to provide a restoring beam of $73\times 26\,$mas and a position angle of $-82.4^\circ$. The rms noise in the image is $54\,\mu$Jy per beam at the effective central frequency of 2276\,MHz. The contours of this image are overlaid on the combined F098M/F105W/F140W image in the right panel of Fig.~\ref{fig:overlay}. We estimate the absolute astrometry uncertainty to be $\le 1\,$mas, and is an equal combination of systematic uncertainty from a 2 degree target-calibrator separation at 2.4\,GHz and the random position uncertainty ($\approx\frac{\rm beamsize}{2\times {\rm SNR}}$). 

\subsection{Literature Data}

\subsubsection{Radio Data}
\label{sec:radio}
D20 presented 25 radio photometric measurements of \gh, including 20 from the GLEAM survey, one from the \ac{nrao} \ac{vla} Sky Survey \citep[\acs{nvss}][]{Condon:98}, and four across $5-20\,$GHz from the \acs{atca} \citep{Frater:92}. This target was originally selected due to its negative curvature across the GLEAM band, $70-230\,$MHz, which suggested a flattening of the spectral index towards the lower end of that range. Modelling of the GLEAM and higher frequency data (see D20, D21 and S22) found that the radio spectrum was best fit by a double power-law with a break frequency of $\approx 2\,$GHz, $\alpha_{\rm LOW}\approx -0.9$ and $\alpha_{\rm HIGH}\approx -1.8$. A triple power-law with a low frequency turn-over below 70\,MHz, to account for the 70-230\,MHz curvature, was found not to provide a better model fit, despite the evident curvature across this frequency range.

\subsubsection{Optical and \ac{ir} data}
\label{subsec:lit}

The Third Data Release of the \ac{hsc} Subaru Strategic Program \citep{Aihara:22} is significantly deeper than the Second Data Release used in S22. \gh~falls in the `Wide' layer of this survey so we downloaded the new, deeper $grizy$ data from the public data release webpage\footnote{\url{hsc-release.mtk.nao.ac.jp/doc/}}. We used aperture photometry on the \ac{wfc3} position of \gh~(Table~\ref{tab:pos}) with the same $0.7\,$arcsec radius aperture and $3-5\,$arcsec annulus as in \S~\ref{subsec:hst} with the aperture corrections determined in S22. As the $y$-band image was still less deep than the F098M image we excluded it from this analysis. As we used a different method for the photometry in this work (compared to D20 and S22) we re-extracted the flux density of \gh~in the HAWKI $K$-band image from D20 using the {\tt photutils} method described in \S~\ref{subsec:hst}. The HSC $griz$ and HAWKI $K$-band photometry, along with that from WFC3, is presented in Table~\ref{tab:hst} and used in the SED fitting presented in \S~\ref{sec:res:beagle}. We also use the mid-\ac{ir} non-detections from the \acl{wise} \citep[\acs{wise},][]{Wright:10} given in S22 to constrain the \ac{agn} bolometric luminosity (\S\ref{sec:dis:mass}). 

\subsubsection{\ac{alma} Continuum Data}
\label{subsec:alma}

We use the \ac{alma} 100\,GHz continuum data from D20. This data comprised five 10-min spectral scans across $85-115\,$GHz. No significant spectral features were found. The 100\,GHz contours overlaid on the F140W image in the left panel of Fig.~\ref{fig:overlay} is a naturally weighted image with an rms of $10\,\mu$Jy per beam, a restoring beam of $1.2''\times 1.4''$, and position angle of $-89^\circ$. The 100\,GHz flux density reported in D21 is $60\pm 13\,\mu$Jy. 

\subsubsection{Interplanetary Scintillation Measurement}
\label{subsec:ips}

The \ac{gama}9 field \citep{driver:11} was covered by \ac{mwa} \ac{ips} observations at 162\,MHz \citep{Morgan:18,Chhetri:18}. \gh~has a \ac{nsi} of $0.49\pm 0.03$ implying that around half the flux is more compact than $\approx 0.3$\,arcsec or there are two compact components separated by more than $\approx 0.3\,$arcsec (see D20 for details). 

\subsubsection{\acs{erosita} Data}

The \acl{erosita} \citep[\acs{erosita};][]{Predehl:21} instrument on the \ac{srg} orbital observatory \citep{sunyaev:21} carried out an early survey of the \ac{gama}9 field, in which \gh~is situated, to the final depth of the planned all-sky surveys. We find that \gh~is undetected in the \acl{efeds} \citep[\acs{efeds},][]{Brunner:22} implying an upper limit to its X-ray $0.5-2.0\,$keV flux of $6.5\times 10^{-15}\,$ erg\,cm$^{-2}$\,s$^{-1}$, the point source $80\%$ completeness limit of \ac{efeds}.

\section{RESULTS AND ANALYSIS}
\label{sec:res}

\subsection{\ac{hst} data}
\label{sec:res:hst}

\subsubsection{Imaging}

Both galaxies are detected in the new F140W image as well as the F105W image. The host of \gh~remains undetected in the F098M image. The F098M and F105W photometry is consistent within 10\% of that reported in S22 with the small difference likely due to slightly different background subtracted method (local here vs global in S22). The left panel of Fig.~\ref{fig:overlay} presents the F140W image overlaid with the 100\,GHz \ac{alma} contours. The host is extended towards the nearby source. The strongest line detected in the host ([OII]$\lambda 3727$ doublet, hereafter [OII] - see \S\ref{sec:res:hst:spec}) is found to be spatially extended by {\tt Grizli} (see Fig.~\ref{fig:spec:host}). Hence, in the right panel we present the combined near-\ac{ir} image overlaid with contours of the [OII] emission map and our \ac{lba} observations (see \S\ref{sec:res:lba}). The [OII] line and \ac{lba} continuum emission are co-spatial and also extended towards the nearby source. However, the spectroscopy discussed in \S\ref{sec:res:hst:spec} confirms that the nearby source is at a lower redshift and that this alignment is purely coincidental. 

\begin{table}[t]
\begin{threeparttable}
\caption{\ac{wfc3}, \ac{hsc} and HAWKI $K$-band aperture corrections, flux densities and uncertainties for the host of \gh. The uncertainties conservatively include an extra 5\% to account for absolute flux calibration and uncertainty in the aperture correction.}
\label{tab:hst}
\begin{tabular}{@{}lcc@{}} \toprule
band & AC & $F_{\nu}$ [$\mu$Jy] \\
\midrule
$g_{\rm HSC}$ & 1.506 & $0.11\pm0.03$  \\
$r_{\rm HSC}$ & 1.703 & $0.31\pm0.06$ \\
$i_{\rm HSC}$ & 1.332 & $0.22\pm0.05$ \\
$z_{\rm HSC}$ & 1.567 & $0.29\pm0.11$ \\
F098M & 1.089 & $0.26\pm 0.14$\\
F105W & 1.093 & $0.50\pm 0.06$\\
F140W & 1.217 & $1.43\pm 0.12$\\
$K_s({\rm HAWKI})$ & 1.327 & $4.12\pm0.37$\\
\bottomrule
\end{tabular}
\end{threeparttable}
\end{table}

\subsubsection{Spectroscopy}
\label{sec:res:hst:spec}

\begin{figure*}[ht!]
\centering
\includegraphics[width=16cm]{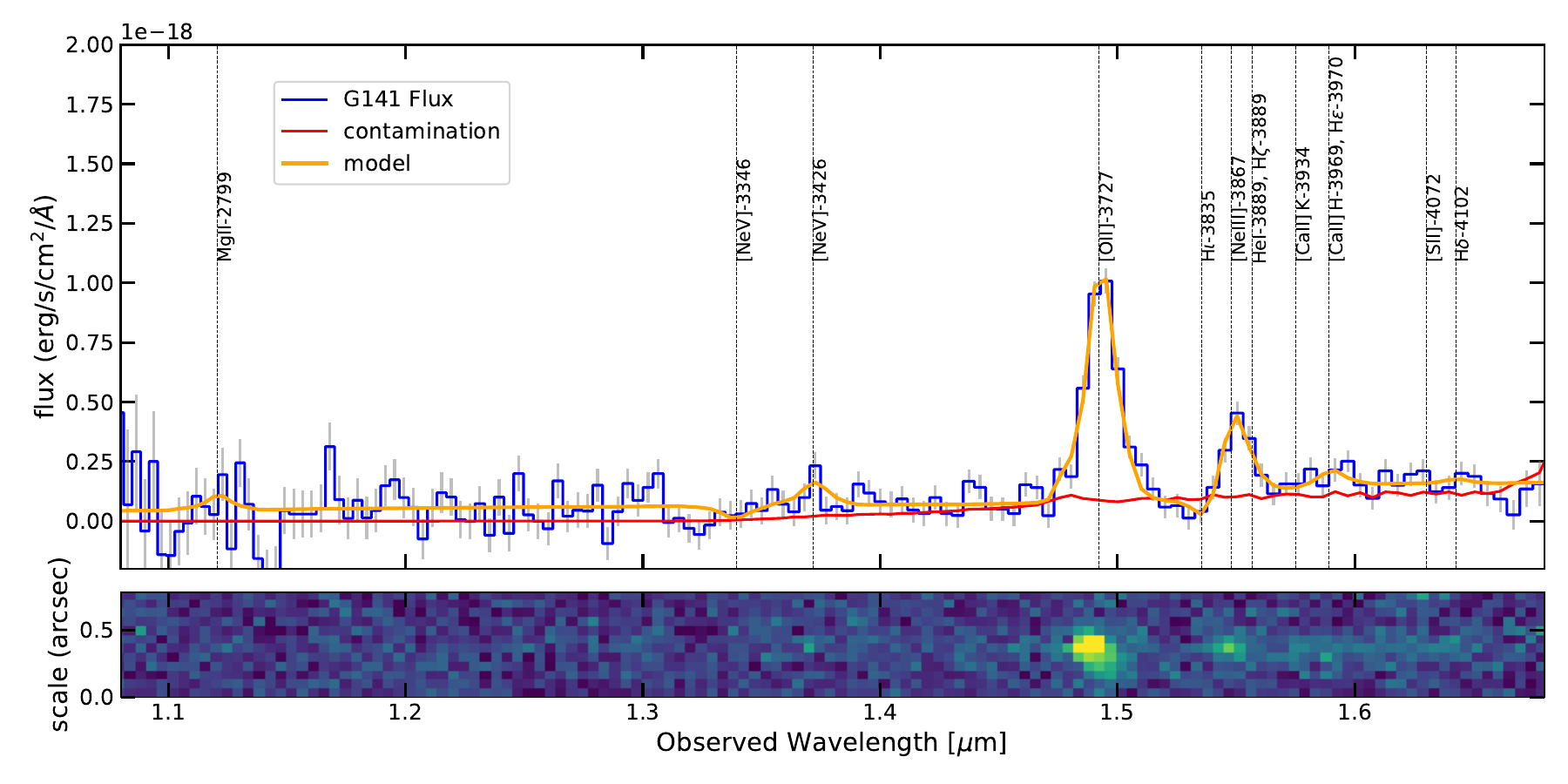}
\caption{\ac{hst}/\ac{wfc3} G141 grism spectrum of the host of \gh~in 1D (top) and 2D (bottom). The \ac{grizli} fitting identifies numerous features, including [OII]$\lambda3727$ (at $\approx 27\sigma$), [NeV]$\lambda3426$ (at $\approx 2.5\sigma$), [NeIII]$\lambda3867$ (at $\approx 7\sigma$), which confirm the redshift of the host of \gh~to be $z=3.00$. The presence of [NeV]$\lambda3426$ and [NeIII]$\lambda3867$ confirm the presence of a high ionisation radiation field from an \ac{agn}. The [OII]$\lambda3727$ is spatially extended as mapped in Fig.~\ref{fig:overlay}.}
\label{fig:spec:host}
\end{figure*}

\begin{figure*}[ht!]
\centering
\includegraphics[width=16cm]{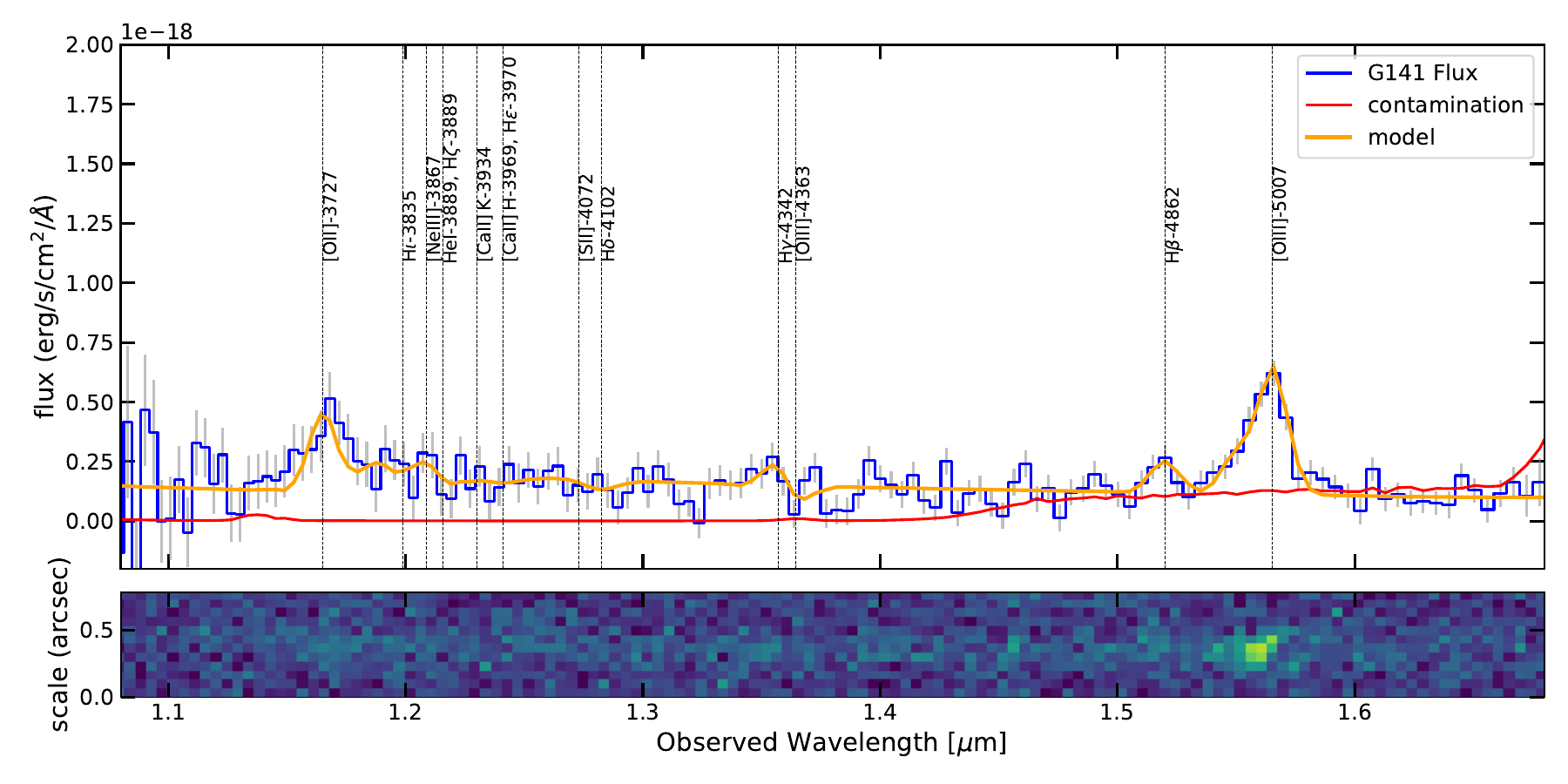}
\caption{\ac{hst}/\ac{wfc3} G141 (blue) grism spectrum of the companion galaxy. The \ac{grizli} software identifies numerous features, including [OII]$\lambda3727$, H$\beta$ and [OIII]$\lambda5007$) which confirm the redshift of the companion galaxy to be $z=2.126\pm 0.001$. Hence, this galaxy is physically unrelated to \gh~which is at $z=3.00$.}
\label{fig:spec:com}
\end{figure*}

In Fig.~\ref{fig:spec:host} we present the G141 spectrum of the host of \gh. The \ac{grizli} software identifies several spectral features, with the strongest line being the [OII] doublet. The other lines {\bf detected} are [NeV]$\lambda3426$ and [NeIII]$\lambda3867$ (hereafter [NeV] and [NeIII] respectively). This combination of lines puts \gh~at $z=3.004\pm 0.001$, implying that the weak line previously identified with the G102 grism (S22) at $1.15\,\mu$m is MgII$\lambda\lambda 2796,2803$.
The properties of the [OII], [NeV] and [NeIII] lines are reported in Table~\ref{tab:hst}.

In Fig~\ref{fig:spec:com} we present the 1D and 2D spectrum for the galaxy $\sim 1.5''$ away from the host of \gh. We also include the best fit {\tt grizli} model to the data which finds the [OII]$\lambda$3727, H$\beta$, [OIII]$\lambda$5007 lines confirming the redshift to be $z=2.126\pm 0.001$, consistent with the photometric redshift of $z_{\rm phot}$$=$$2.2^{+0.3}_{-0.6}$ presented in D21. Hence this source is physically unrelated to the host of \gh and we consider it no further. 

The redshift solution of $z=3.004$ is consistent with the lower of the two options put forward in S22. While this radio galaxy is not at $z>7$, it still demonstrates several unique properties. From the radio photometry presented in D20 and the modelling in \cite{Broderick:22} the 500\,MHz rest-frame luminosity of this source is $L_{\rm 500MHz}=1.0\pm0.1\times 10^{28}\,$W\,Hz$^{-1}$ (conservatively including a 10\% uncertainty from the modelling), putting it well into the `powerful' \acl{hzrg} (\acs{hzrg}: $z>1$ and $L_{\rm 500MHz}\ge 10^{27.5}\,$W\,Hz$^{-1}$) regime \citep[][]{Miley:08,DeBreuck:10}. This result, as well as the 1.4\,GHz and 178\,MHz luminosities, is presented in Table~\ref{tab:sum_radio}. We discuss how this \ac{hzrg} compares to others at the same redshift in \S\ref{sec:dis:comp}.

The [NeIII] and [NeV] lines are  strong indicators of the presence of an AGN \citep{gilli:10} due to their high ionisation potentials \citep[e.g.][]{Maddox:18} and equivalent widths of $\ge 4\,$\AA~are enough to confirm the presence of an obscured \ac{agn}. The [OII] line is typically indicative of star formation and often used as \ac{sfr} tracer \citep[e.g.][]{Kennicutt:92}. However, from the high radio luminosity and neon lines, we know that there is a powerful AGN present. Due to its brightness we can map the [OII] (Fig.~\ref{fig:overlay}) and see that it extends $\approx 0.6''$, equivalent to 4.7\,kpc. Indeed, in the 2D spectrum (Fig.~\ref{fig:spec:host}) one can see the spatially extended [OII]. The [NeV] and [NeIII] lines are too faint to be mapped out. 

Hence, if the [OII] emission is primarily powered by an AGN then we can determine an upper limit to the unobscured \ac{sfr} using the equation 4 of \citet[][K04]{kewley:04} and the relation in \citet[][K98]{Kennicutt:98}. Note no correction is made to the [OII] line luminosity for reddening as the SED fitting in \S~\ref{sec:res:beagle} cannot constrain the dust attenuation well and the three lines are close together. From a line luminosity of $L_{\rm [OII]}=3.28\times 10^{42}$erg\,s$^{-1}$ we obtain a \ac{sfr} upper limit of $<$$22$ and $<$$46\,$M$_\odot$\,yr$^{-1}$ for the K04 and K98 conversions, respectively (see Table~\ref{tab:sum_opt}). 

Powerful radio galaxies have been classically separated into two distinct classes based on the strength of their optical emission lines \citep{Hines:1979}. Radio galaxies with strong emission lines are referred to as high excitation (radio) galaxies, HE(R)Gs, and those with weak emission lines as low excitation (radio) galaxies, LE(R)Gs. The original work was based on the [OII] line strength \citep{Hines:1979}, but newer criteria exist based on a combination of lines \citep[e.g.,][]{Buttiglione:10}. The grism spectra do not cover all the lines to make a formal classification with the \cite{Buttiglione:10} criteria so we make the assumption that the [OIII]$\lambda$5007 line luminosity is within an order of magnitude of the [OII] line luminosity (as is typically seen in radio galaxies). Taking the 178\,MHz luminosity along with the [OIII]$\lambda$5007 line luminosity we can place \gh~on the scatter plot of these two luminosity in Fig. 8 of \cite{Buttiglione:10}. We find that \gh~lies at the high luminosity end of the LERG track and well away from the HERG track. So despite these high luminosities we consider \gh~to be a LERG.

\subsection{LBA Results}
\label{sec:res:lba}

In Fig.~\ref{fig:overlay} we show the contours of the LBA observations over the combined F140W/F105W/F098M image. The relative uncertainty in the astrometry between \ac{lba} and \ac{lba}  is small: the \ac{hst} observations are tied to the {\it GAIA} DR3 reference frame, which is consistent with the \ac{icrf} used by \ac{lba} within $<$$0.01\,$mas \citep{Klioner:22}. This difference is far less than the absolute positional uncertainty of \ac{lba}, $<$$1\,$mas.

The \ac{lba} data show two unresolved components separated by 620\,mas, equivalent to a projected separation of 4.8\,kpc. The \ac{ne} component is closer to the galaxy centre, as inferred from the optical image, but it is far enough away that it is unlikely to be the core. Hence we are likely seeing the compact ends of two jets from the central black hole rather than a binary or offset \ac{agn}. The \ac{sw} component is 470\,mas away from the peak of the optical emission. Assuming the central black hole is located at this optical peak, the projected length of the jet to the SW component is 3.6\,kpc. The flux densities were measured with the {\tt miriad} task {\tt imfit} and found to be $10.4\pm0.4$ and $6.8\pm0.5\,$mJy (not including the 10\% overall flux calibration uncertainty) for the \ac{ne} and \ac{sw} components, respectively. The sum of these fluxes is $\approx 80\%$ of the estimated 2276\,MHz total flux, $S_{\rm 2276MHz}=21.8\,$mJy, from the broad-band modelling presented in \cite{Broderick:22}. Hence, $\approx 20\%$ of the flux density may be resolved out in the \ac{lba} observations. If these two components indicate the ends of the jets/lobes and if we assume that the black hole is equidistant to the two hot spots, then we estimate an age of $\ge 78\,$kyr assuming the jet propagates at $0.1\,c$. This propagation speed is a median value for \ac{gps}/\ac{css} sources in the review by \cite{odea:21}. Faster speeds, as assumed by the jet modelling in \S~\ref{sec:res:jet} would imply younger ages. This estimate is also consistent with the slowing jet propagation speeds for older jets seen in powerful GPS sources \citep{An:12}.

The compact size of the radio emission is $\ll 20\,$kpc, consistent with the radio source being classified as CSS. The extended [OII] is aligned with the SW radio component, which is a common property of CSS sources \citep{devries:97}. We discuss the extended [OII] further in \S\ref{sec:dis:comp}. The observed asymmetry suggests that the NE component is likely impacting denser ISM than the SW component. 

The two compact component model with some extended flux is also consistent with the morphological models put forward to explain the IPS value of NSI $=0.49\pm 0.03$ reported in D21. An NSI of unity implies an unresolved source at 162\,MHz (i.e. $\leq 0.3\,$arcsec) whereas a value of zero implies all the flux is extended on larger scales. An in-between value implies either that only a fraction of the flux is compact and/or that there is more than one compact component, as is the case here. 

\subsection{Modelling the Host Galaxy}
\subsubsection{SED}
\label{sec:res:beagle}

We use the SED fitting code \acl{beagle} \citep[\acs{beagle};][]{Chevallard:16}\footnote{\url{www.iap.fr/beagle/}} to constrain the properties of the host galaxy. For this fitting we use the six photometric measurements in Table~\ref{tab:hst}, including three from WFC3, four from HSC and the one from HAWKI $K$-band. Inclusion of $<3\sigma$ {\it WISE} limits from S22 made no appreciable difference to the host galaxy properties nor their uncertainties. 

Within \acs{beagle} we use a delayed exponential \ac{sf} history, a \cite{Calzetti:97} dust attenuation, and broad priors. The ionisation parameter and metallicity of the nebular regions is set equal to that of the stars. The dust to metal mass ratio is left as default value of 0.3. Hence, we allow only the following five parameters to vary: timescale of \ac{sf} (hereafter `age'), stellar mass, current SFR (we report the SFR averaged of the last 100\,Myr), metallicity and extinction.

\begin{figure}[t]
\centering
\includegraphics[width=8.5cm]{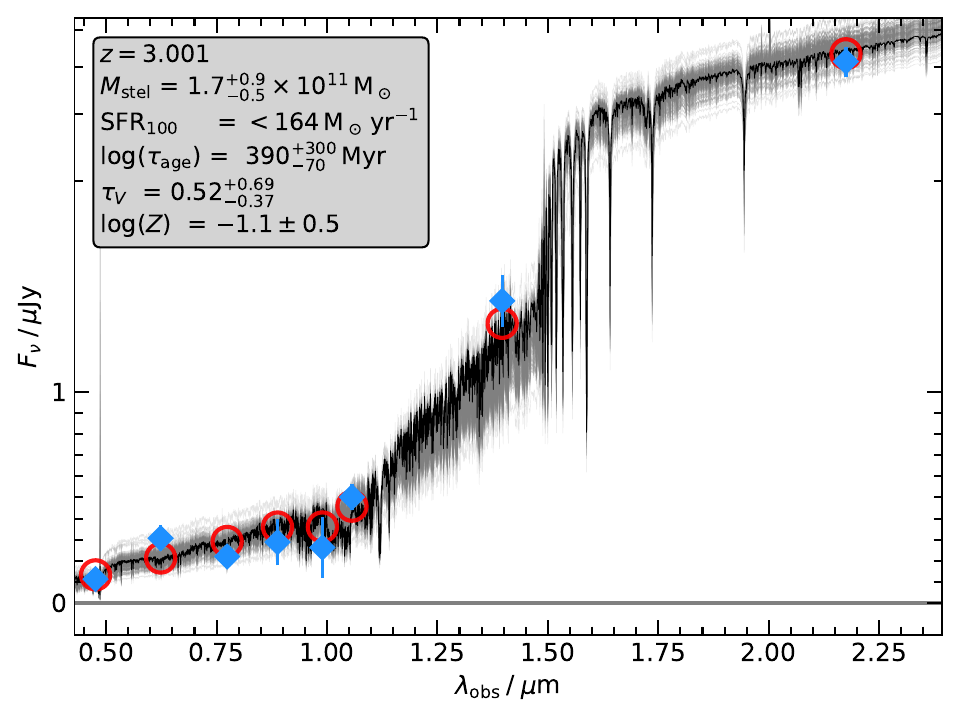}
\caption{Fit of the optical/near-IR photometry of \gh~with \acs{beagle} showing the five photometric detections (blue diamonds) and uncertainties (sometimes obscured by the symbols). Also shown is the template with the highest probability (in black) as well as all the fits within all the $68\%$ confidence limits (in grey). The values of the photometry from the highest probability model are shown in red. The median SED parameters (or limits) for this fit are presented in Table~\ref{tab:sum_opt}.}
\label{fig:beagle}
\end{figure}

\acs{beagle} provides PDFs from which we take the median value as the best fit and the 68\% confidence intervals for the uncertainties. In the case where the difference between the upper and lower 68\% confidence interval is more than an order of magnitude we take the 95\% confidence upper value as a $2\sigma$ upper limit. This affects only the SFR. These results are reported in Table~\ref{tab:sum_opt} and the highest probability SED template is plotted in Fig~\ref{fig:beagle}. 

Examining the bivariate PDFs (i.e. `corner plots') there is some slight degeneracy between extinction and the stellar mass and age. However, the 68\% confidence interval on the stellar mass ($1.7^{+0.9}_{-0.5}\times 10^{11}\,$M$_\odot$) and age ($39^{+300}_{-70}\,$Myr) are well constrained. The SFR is poorly constrained with the 68\% confidence interval spanning three orders of magnitude. We thus report the 95\% confidence interval upper limit for the SFR, but we note this is dominated by a secondary PDF peak corresponding to a very unobscured ($\tau_V\sim 1$) and very young ($\tau\sim 10\,$Myr) starburst. Hence, the true upper limit to the SFR is likely a lot lower ($\ll 10\,$M$_\odot$ from visual inspection of the PDF around the principle PDF peak).

\subsubsection{Morphology}
\label{sec:res:petro}
We use the F140W image to fit a \cite{Sersic:68} profile to the host galaxy using {\tt PetroFit} \citep{Geda:22}. The cutout fed to {\tt PetroFit} was $3\times 3\,$arcsec and the \ac{rms} image was created at a constant value equal to that of the locally determined off-source \ac{rms}. The S\'ersic model was convolved with the F140W PSF\footnote{\url{https://www.stsci.edu/hst/instrumentation/wfc3/data-analysis/psf}} which was over-sampled by a factor of two relative to the drizzled pixel size. We find an effective radius (i.e. half-light radius), $R_{\rm eff}$, of $2\pm0.3\,$kpc and a S\'ersic index of $n= 3.3 \pm 0.6$. We also determine $R_{\rm 80}$, corresponding to the radius encompassing 80\% of the total flux \citep[using the expression from ][]{Miller:19}, which has the advantages of: (a) encompassing a larger fraction of the baryons and (b) finding similar sizes for both \ac{sfgs} and quiescent galaxies \citep{Mowla:19}. Using this relationship, which is a function of $R_{\rm eff}$ and the S\'ersic index, we obtain $R_{\rm 80}=5.8\pm 1.1\,$kpc where the uncertainty is propagated from the inputs. These results are presented in Table~\ref{tab:sum_opt}, discussed in \S\ref{sec:dis:host} and are used in modelling the radio jet in \S\ref{sec:res:jet}.

\subsection{Modelling the Radio Jet}
\label{sec:res:jet}

\begin{figure*}[th!]
\centering
\includegraphics[width=16.5cm]{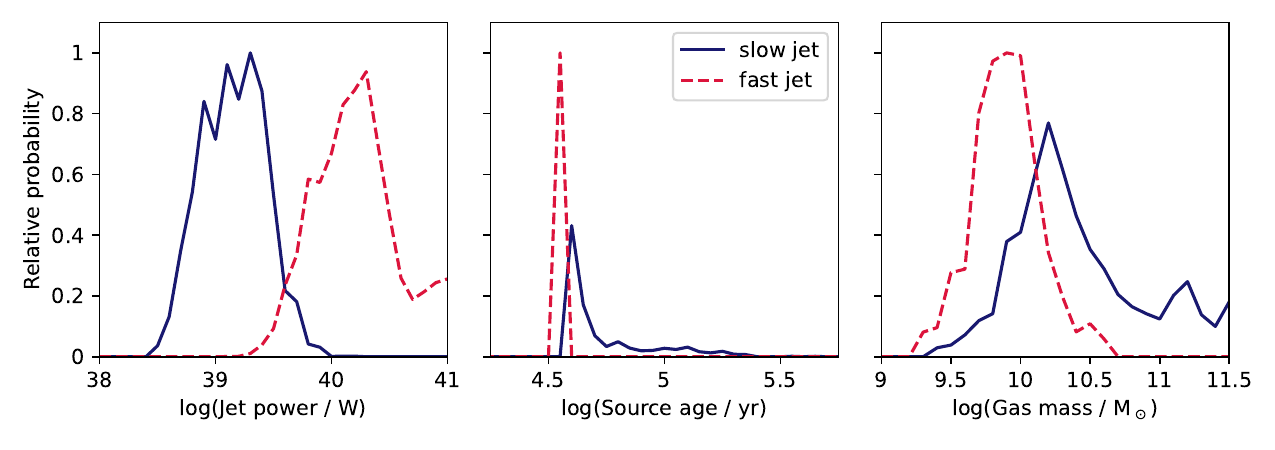}
\caption{The probability distribution functions of the three free parameters from the \ac{raise} modelling of the radio jet (see \S\ref{sec:res:jet}). Two potential solutions are found: a `fast' jet (red dotted line - higher jet power) and a `slow' jet (blue solid line - lower jet power). As discussed in the text, we favour the `slow' jet solution whose best fit results are presented in Table~\ref{tab:sum_radio}.}
\label{fig:raise}
\end{figure*}

We constrain the jet power and age of the radio jet in \gh~using a parameter inversion informed by simulations from the latest version of \acl{raise} \citep[\acs{raise},][]{Turner:15,Turner_2023}. The compact size of this source suggests its evolution will be dominated by the ISM of the host galaxy; we therefore model the distribution and density of this ambient medium in \ac{raise} using Monte Carlo realisations of the effective radius and S\'ersic profile fitted in Section \ref{sec:res:petro}. In doing this we are assuming that the ISM has the same distribution in the galaxy as the stars. 

The expansion rate (velocity) of the jet-head is proportional to the velocity of the bulk flow in the jet spine (Lorentz factor of $\gamma_\text{j} =5$), with a constant depending on the relative densities of the jet, $\rho_\text{j}$ and the ISM cloud, $\rho_\text{c}$, it is propagating through \citep[see equation 5 from][]{Turner_2023}. Hence, the travel time through the ISM, i.e. age of the jet, is simply the integral of the inverse of the jet-head expansion rate along the jet path. This integral can include any variation in the ISM density, but for a fixed overall gas mass it is independent of the density profile. 

Local density inhomogeneities in the multiphase ISM will not affect the jet expansion -- compared to a smooth profile with the same total mass -- provided no very dense clouds (i.e., $\rho_\text{c} \gg \gamma_\text{j}^2 \rho_\text{j}$; \citealt{Turner_2023}) lie in the path of the jet. We therefore assume the density is locally homogeneous but note as the radio source is asymmetric a slightly older source age is possible if a dense cloud is encountered (Young et al., {\it in prep.}). The total mass of gas encountered by the jet remains a free parameter; we therefore constrain the total gas mass (the sum of the molecular, neutral and ionised gas, i.e. the sum of Baryonic matter minus the stars) as a third free parameter (in addition to the jet power and age).

\ac{raise} model outputs (synthetic images and SEDs) are generated over a three-dimensional grid covering all reasonable values of these three parameters. We assume the default values for other parameters in \ac{raise}, noting these are calibrated against hydrodynamical simulations and/or based on observations of well-studied sources. The equipartition factor (ratio of energy in the magnetic field to energy in the particles) assumes a prior probability density function following \citet{Turner:18b}. 

These synthetic images at 1.4\,GHz are compared to the following observable values using a maximum likelihood approach: NVSS flux density, and the frequency-invariant extent of the jets from the LBA image. We find the steep two-point spectral index between 1.4\,GHz and 5.5\,GHz (photometry from NVSS and the ATCA, respectively) can be explained if the freshly shock-accelerated electron population is at a significantly lower magnetic field than the remainder of the population, as occurs when the jet-head passes the half-light radius of the galaxy, $R_\text{eff}$, into the steep, outer-sections of the galaxy S\'ersic profile. This is consistent with our constraint on the extent of the jets. We do not include lower  frequencies in the modelling as RAiSE currently does not include free-free absorption, a likely cause of the radio spectral flattening.

We find two potential solutions shown in the PDFs presented in Fig.~\ref{fig:raise}: a `fast' jet with a (slightly) younger age and higher jet power, i.e. a source that expands at the speed of the bulk flow along the jet (hence the small scatter in jet ages) and a `slow' jet with an older age and lower jet power which has at least some interaction with the ambient medium leading to a slowing of the jet. The `slow' solution has a higher gas mass since the jet has to work harder to expand. The gas could be slightly more extended than inferred from the stellar morphology which would decrease the jet power and increase the source age. Conversely if the gas is slightly less extended then the jet power would increase and the gas mass decrease. In \acs{hzrgs} we tend to see denser gas on more extended scales, often along the jet axis \citep{Klamer:09,Emonts:14}. Of the two solutions, we believe the `slow' jet to be more likely. First, the jet power distribution is very steep at the high powers \citep{Quici:24} 
so a random source is more likely to be at lower luminosity. Second, the hosts of \acs{hzrgs} are more likely to be gas rich \citep{Emonts:14}. 

The best fit result and uncertainties for the slow jet solution are presented in Table~\ref{tab:sum_radio} summerised as a jet power of $Q_{\rm jet}=1.5^{+1.3}_{-0.7} \times 10^{39}\,$W,  an age of $47^{+43}_{-6}\,$kyr (n.b. the peak in the age PDF will shift with the assumed bulk flow velocity) and a gas mass of $2.0^{+5.9}_{-1.1}\times 10^{10}\,$M$_\odot$. 

\begin{table}[t]
\begin{threeparttable}
\caption{Summary of the optical to near-IR properties of the host galaxy of \gh}
\label{tab:sum_opt}
\begin{tabular}{@{}lc@{}} 
\toprule
parameter & value \\
\midrule
\multicolumn{2}{c}{WFC3/\ac{grizli}}\\
\hline
redshift, $z$ & $3.004\pm0.001$  \\
lines observed &  [OII]$\lambda\lambda3727$, [NeV]$\lambda3426$,  [NeIII]$\lambda3867$ \\
$S_{\rm [OII]}$ &  $1.61\pm 0.06 \times 10^{-16}\,$erg\,s$^{-1}$\,cm$^{-2}$\\
$L_{\rm [OII]}$ &  $3.28\pm 0.12 \times 10^{42}\,$ erg\,s$^{-1}$\\
$S_{\rm [NeV]}$ &  $1.52\pm 0.6 \times 10^{-17}\,$erg\,s$^{-1}$\,cm$^{-2}$\\
EW$_{\rm [NeV]}$ & $224\,$\AA\\
$L_{\rm [NeV]}$ &  $3.1\pm 1.2 \times 10^{41}\,$ erg\,s$^{-1}$\\
$S_{\rm [NeIII]}$ &  $5.42\pm 0.77 \times 10^{-17}\,$erg\,s$^{-1}$\,cm$^{-2}$\\
$L_{\rm [NeIII]}$ &  $1.10\pm 0.16 \times 10^{42}\,$ erg\,s$^{-1}$\\
FWHM$_{\rm [OII]}$ (obs.) & $157\pm 8\,$\AA \\
FWHM$_{\rm [OII]}$ (rest) & $<3160\pm 160\,$km\,s$^{-1}$ \\
SFR$_{\rm [OII]}$ (K04) & $<22\,$M$_\odot$\,yr$^{-1}$  \\ 
SFR$_{\rm [OII]}$ (K98) & $<46\,$M$_\odot$\,yr$^{-1}$  \\ 
\midrule
\multicolumn{2}{c}{\acs{beagle} fitting results}\\
\midrule
stellar mass, $M_{\rm stel}$ & $1.7^{+0.9}_{-0.5}\times 10^{11}\,$M$_\odot$   \\
SFR$_{\rm 100}$     & $<164\,$M$_\odot$\,yr$^{-1}$  \\
$\log(\tau_{\rm age})$ &  $390^{+300}_{-70}\,$Myr\\
$\tau_V$ & $0.52^{+0.69}_{-0.37}$\\
$\log(Z)$ & $-1.1\pm 0.5$\\
\midrule
\multicolumn{2}{c}{{\tt PetroFit} fitting results}\\
\midrule
$R_{\rm eff}$ & $2.0\pm0.3\,$kpc\\
S\'ersic index, $n$ & $3.3 \pm 0.6$\\
$R_{\rm 80}$  & $5.8\pm1.1\,$kpc\\
\midrule
\multicolumn{2}{c}{other data}\\
\midrule
sSFR$^a$ & 0.23\,Gyr$^{-1}$\\
$S_{\rm 0.5-2keV}$ (eROSITA) & $<6.7\times 10^{-15}\,$ erg\,s$^{-1}$\,cm$^{-2}$  \\
$L_{\rm 0.5-2keV}$ (eROSITA)$^b$ & $<1.7\times 10^{44}\,$erg\,s$^{-1}$\\
$L_{\rm BOL}$(\rm [NeV]) & $1.1\pm 0.4\times 10^{12}\,$L$_\odot$ \\
$L_{\rm BOL}$(\rm mid-IR) & $<5\times 10^{12}\,$L$_\odot$\\
\bottomrule
\end{tabular}
\end{threeparttable}
    \begin{tablenotes}
      \small
      \item $^a$derived from the higher of the [OII] SFR upper limits. $^b$ assuming a photon index of $\Gamma=1.8$.\\
    \end{tablenotes}
\end{table}

\begin{table}[t]
\begin{threeparttable}
\caption{Summary of radio and derived properties of \gh.}
\label{tab:sum_radio}
\begin{tabular}{@{}lcc@{}} \toprule
parameter & value\\
\midrule
\multicolumn{2}{c}{Observables}\\
\midrule
{\bf $178\,$MHz luminosity, $L_{\rm 178MHz}$} & {\bf $1.7\pm0.2\times 10^{28}\,$W\,Hz$^{-1}$}  \\

$500\,$MHz luminosity, $L_{\rm 500MHz}$ & $1.1\pm0.1\times 10^{28}\,$W\,Hz$^{-1}$  \\
$1.4\,$GHz luminosity, $L_{\rm 1.4GHz}$ & $5.2\pm0.5\times 10^{27}\,$W\,Hz$^{-1}$ \\
Largest angular extent & 4.8\,kpc\\
\midrule
\multicolumn{2}{c}{\ac{raise} (slow jet)}\\
\midrule
Jet power, $Q_{\rm jet}^{\tt RAiSE}$ & $1.5^{+1.3}_{-0.7}\times 10^{39}\,$W  \\
Jet age, $t_{\rm jet}$ & $47^{+43}_{-6}\,$kyr \\
gas mass, $M_{\rm gas}$ & $2.0^{+5.9}_{-1.1}\times 10^{10}\,$M$_\odot$ \\
\midrule
\multicolumn{2}{c}{Derived}\\
\midrule
black hole mass, $M_{\rm BH}$ & $\ge 10^{9}\,$M$_\odot$   \\
$\lambda_{\rm EDD}$ &  $\le 0.03$ \\
\bottomrule
\end{tabular}
\end{threeparttable}
\end{table}

\section{DISCUSSION}
\label{sec:dis}

The measured and derived optical/near-IR and radio properties of \gh~are presented in Tables~\ref{tab:sum_opt} and~\ref{tab:sum_radio} respectively.

\subsection{Host Galaxy Properties}
\label{sec:dis:host}
The stellar mass and size of this galaxy as measured by $R_{\rm eff}$ put \gh~around a factor of two below the mean mass-size relation in the redshift bin $2.5<z<3.0$ as determined from deep {\it HST} surveys of field galaxies \citep{Vanderwel:14}. Using $R_{\rm 80}$, which is arguably a better measure of galaxy size (see \S\ref{sec:res:petro}), \gh~is still a factor of 1.5 below the mean mass-size relation for the same redshift bin \citep{Mowla:19}. Even allowing for a slight evolution with redshift, as well as the paucity of massive galaxies in pencil-beam {\it HST} surveys, this result suggests that \gh~is at the lower edge of the distribution of field galaxies, i.e. compact for its stellar mass and redshift. 

The stellar mass, $M_{\rm stel}\approx 2\times 10^{11}\,$M$_\odot$, puts \gh~at the low end of distributions of stellar mass seen in HzRGs \citep{Seymour:07,DeBreuck:10} perhaps consistent with it being at the low end of the radio luminosity distribution for this sample. This stellar mass is consistent with the total gas mass, $M_{\rm gas}\approx 2\times 10^{10}\,$M$_\odot$, derived in the \acs{raise} model fitting of the jet: $M_{\rm gas}/M_{\rm stel}\approx 0.1$. The SFR is poorly constrained by  \acs{beagle} due to the wavelength coverage of available photometry, but the limits from the [OII] emission ($<46\,$M$_\odot\,$yr$^{-1}$) infer a specific SFR (sSFR) of 0.23\,Gyr$^{-1}$. This value puts \gh~at least 0.5\,dex below the main sequence for star forming galaxies at this redshift \citep{Koprowski:24}.

As the CO(4$\rightarrow$3) line is not detected at 86.4\,GHz in the 85-115\,GHz ALMA spectral scan reported in D21, we infer a limit to the CO luminosity of $L'_{\rm CO}< 10^{10}\,$K\,km\,s$^{-1}$\,pc$^2$. This luminosity is equivalent to a limit on the molecular gas mass of $M_{\rm mol}<4.3\times 10^{10}\,$M$_\odot$. The implied molecular gas fraction, $f_{\rm gas}\equiv\frac{M_{\rm mol}}{M_{\rm stel}+M_{\rm mol}}$, is very low, $f_{\rm gas}\le 0.09 $, consistent with the lower gas fractions seen in more massive galaxies at this redshift (e.g. P\'erez-Mart\'inez et al., submitted) with the caveat that few CSS sources have well-measured gas fractions \citep{odea:21}.

\subsection{Comparison to Other HzRGs and GPS/{\bf CSS} Sources}
\label{sec:dis:comp}

\begin{figure}[t]
\centering
\includegraphics[width=8.5cm]{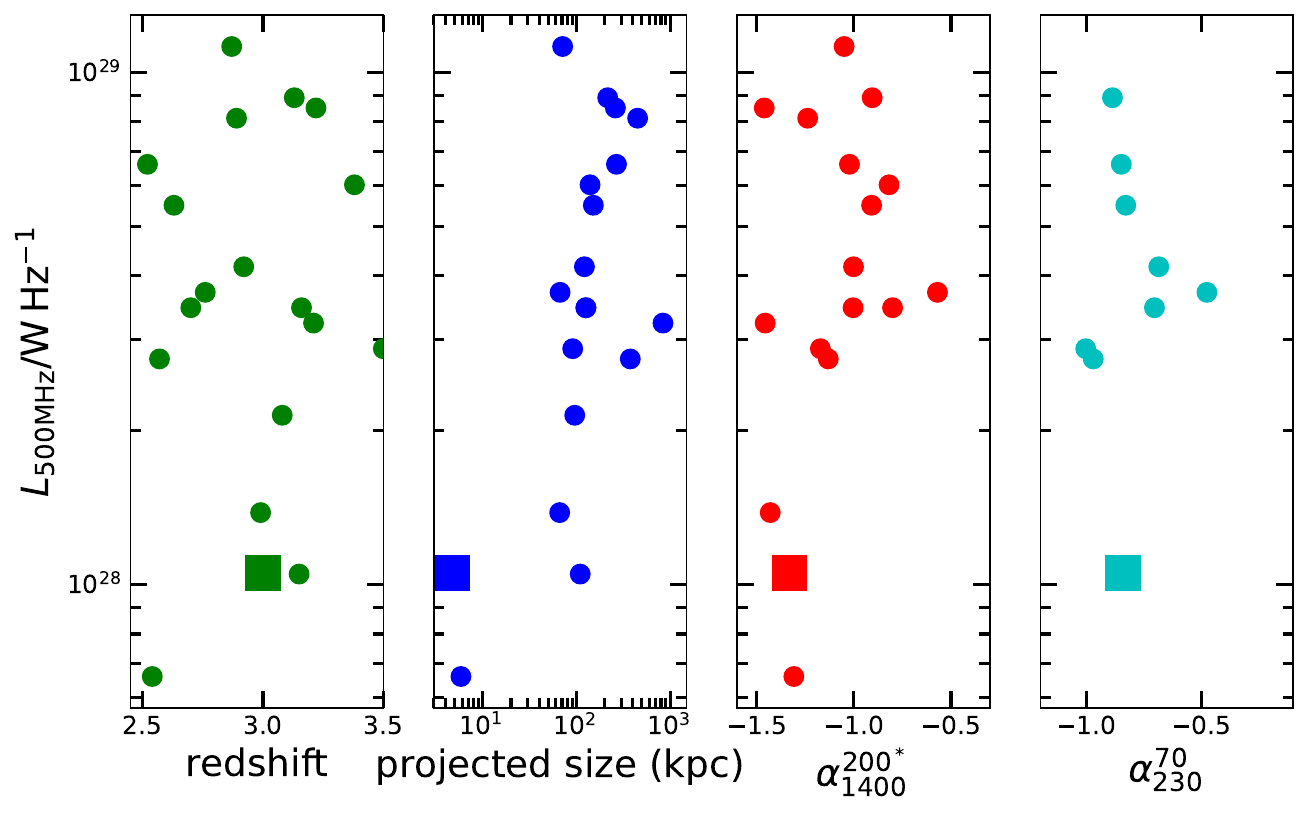}
\caption{Properties of \gh~(square) compared to powerful HzRGs (circles) from \cite{Seymour:07} and \cite{DeBreuck:10}. Left to right shows 500\,MHz luminosity plotted as a function of redshift, projected angular size, and the spectral index measured at high frequency ($151/187/200/325-1400\,$MHz) and low frequency ($70-230\,$MHz). In the final two panels we omit sources lacking low-frequency data and those not covered by GLEAM.}
\label{fig:quadtic}
\end{figure}

The fitting with \ac{raise} (\S\ref{sec:res:jet}), unsurprising for such a compact source, finds a very young radio galaxy $\approx 50\,$kyr, but with a very high jet power. Note, it is unlikely that this radio source is beamed given the steep spectrum ($-1.7<\alpha<-0.8$ across $0.1<\nu_{\rm observed}/{\rm GHz}<30$), the $<2.6\%$ linear polarisation at 7.25\,GHz (D21), and the fact we see both jets.

Fig.~\ref{fig:quadtic} compares the properties of \gh~to the sample of powerful HzRGs reported in \cite{Seymour:07} and \cite{DeBreuck:10}. We limit the comparison to $2<z<4$ and compare 500\,MHz luminosities to the redshift, size, and high frequency (151/187/200/325-1400\,MHz) and low frequency (70-230\,MHz) spectral indices. The latter is only available for the GLEAM detected sources and the higher frequency spectral indices are derived from NVSS \citep{Condon:98}, i.e. 1.4\,GHz, and GLEAM 200\,MHz flux densities; where GLEAM data was not available, we use other low-frequency photometry: 178\,MHz -- \citep{Pilkington:65}, 151\,MHz -- \citep{Waldram:96} and 325\,MHz -- \cite{Rengelink:97}. \gh~is on the lower end of the luminosity distribution and is smaller than all sources bar WN J1115$+$5016 (at $z=2.54$) which is a compact {\bf radio} double with a size of $0.2''$ (1.64\,kpc). WN J1115$+$5016 shows CIV$\lambda$1549 in broad absorption \citep{DeBreuck:01} suggesting an outflow like other broad absorption line QSOs. 

\gh~also has a steep high frequency spectral index compared to most HzRGs in this sample. Modelling of high redshift USS-selected radio galaxies suggests that the steep spectrum is partly by selection and partly due to higher inverse Compton losses at higher redshifts \citep{Morabito:18}. As this source was selected at low-frequency, its USS nature is not by selection. Its steep high-frequency spectrum must be due to inverse Compton or simple synchrotron losses. The modelling from \acs{raise} in \S\ref{sec:res:jet} suggests that the steepening is likely caused by synchrotron losses combined with the sudden drop in magnetic field strength experienced by freshly shock-accelerated electrons upon exiting the galaxy.

Comparing the age and size of \gh~to a sample of bright CSO sources across $2<z<2.37$ \citep{murgia:99}, we find that it lies long the track of lobe dominated, rather than jet dominated, CSO sources. This result is not surprising as most ($\approx 80$\%) of the radio emission is detected on LBA scales  and our modelling with \acs{raise} in \S\ref{sec:res:jet} is based on these (small) lobes.

The extended component of the [OII] line in \gh~is reminiscent of the jet-cloud interactions observed in low-redshift sources like 3C316 \citep{An:13} and 3C48 \citep{Stockton:07,An:10}. In these objects, the jet morphology shows clear signatures of strong interactions, including knotty jet structures with bright jet knots and large jet bending indicative of deflection. The extension of the [OII] emission along the jet axis of GLEAM J0917–0012 is also reminiscent of alignments seen between the radio plasma and various constituents of the host environment in extended HzRGs, which often also show complex and knotty radio morphologies \citep[][see also Sect. 1]{Miley:08}. 

From the broad-band radio spectral fitting in D21, the peak flux must be at $\ll 280\,$MHz (in the rest-frame) which potentially puts this source on or below the intrinsic turnover frequency vs. linear size relation presented in \cite{odea:97}. We note that the most comparable source to \gh, WN J1115$+$5016 (discussed above), must also have a rest-frame peak frequency $<255\,$MHz and hence may be another high redshift MPS/CSS source.

The X-ray non-detection with eROSITA implies a luminosity lower limit of $L_{\rm 0.5-2keV}<1.7\times 10^{44}\,$erg\,s$^{-1}$ (see Table\ref{tab:sum_opt}) assuming a photon index of $\Gamma=1.8$. Comparing this result to other X-ray observations of CSS sources \citep{kb:14} we see that (a) \gh~lies in the region of parameter space occupied by FR II radio galaxies \citep{fr:74} and (b) it could be consistent with either a HERG or LERG. The former result is not too surprising as the LBA image detects the majority of the flux finding two compact lobes. As we determined in \S\ref{sec:res:hst:spec} that this source is a LERG then we would expect the true X-ray luminosity to be roughly an order of magnitude below the limit \citep[see Fig. 3 of ][]{kb:14}.

\subsection{Origin of Ionising Field}
\label{sec:dis:ion}

As discussed in \S\ref{sec:res:hst:spec}, the presence of [NeV] is a strong indicator of an AGN and is often seen in radio galaxies with EELRs \citep[e.g.][]{Spinrad:84}. Compact radio galaxies with EELRs aligned with the radio emission are generally powered by shock ionisation \citep{DeBreuck:00,Fu:09a} as supported by observations of UV line ratios. Studies of EELRs typically focus on the [OIII]$\lambda$5008 and H$\alpha$ lines since the lower critical density of [OII] means it can arise well away from the narrow-line region in the EELR \citep[e.g.][]{VillarMartin:11a}. 

Looking at the bright [OII] emission from \gh~we estimate an upper limit to the velocity dispersion by fitting a single Gaussian to the 1D spectrum finding ($\equiv 3160\pm 160\,$km\,s$^{-1}$). This will  over-estimate the velocity width as source morphology and velocity are convolved in grism observations (hence it is reported as an upper limit in Table~\ref{tab:sum_opt}). Higher resolution prism/slit spectroscopy is required to determine an accurate value. 

\begin{figure*}[t]
\centering
\includegraphics[width=14cm]{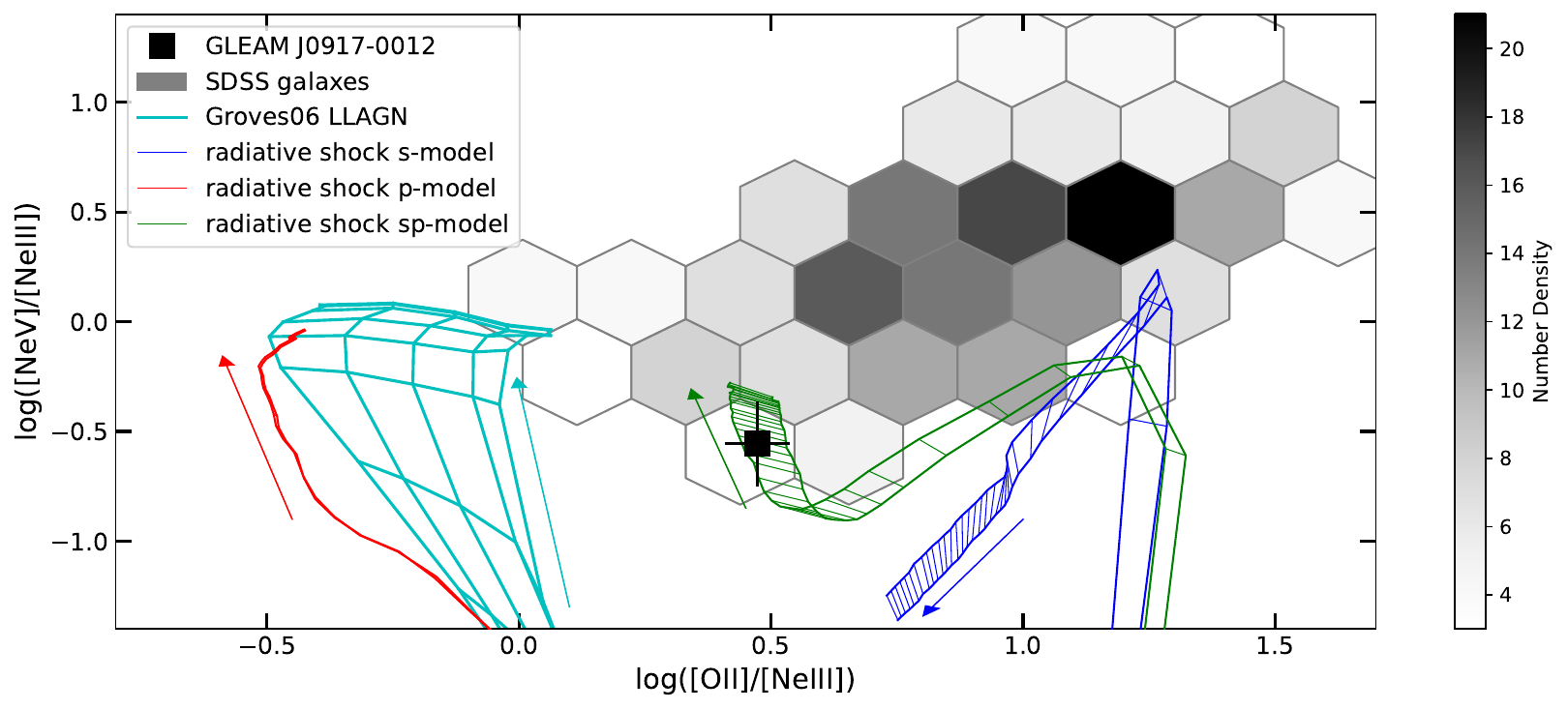}
\caption{Observed [NeV]/[NeIII] vs [OII]/[NeIII] emission line flux ratios of \gh~(square) compared to models and field galaxies from SDSS (limited to [NeV] flux $\ge 2\sigma$ and [NeII], [OII] fluxes $\ge 2.5\sigma$). The field galaxies are a mix of SFGs and AGN according to classic BPT analysis with the density bins only shown when number of galaxies $N\ge 3$. We also show the MAPPINGS III \citep{Allen:08} tracks of the radiative shock models for a galaxy with a low, \ac{smc}-like metallicity (comparable to the host, see Table~\ref{tab:sum_opt}). The three different models are radiative shock and photoionised precursor (sp) models and the simpler shock (s) or precursor (p) only models (with the arrows indicating the direction of increasing shock velocity from 100 to 1,000\,km\,s$^{-1}$). Finally, we show the tracks of \ac{llagn} from \cite{Groves:06} with different metallicities and ionisation parameters (the arrow indicates the direction of increasing ionisation parameter value). From this analysis the most likely cause of the line ratios seen in \gh~is a radiative shock with a photoionised precursor.}
\label{fig:lineratio}
\end{figure*}

Fig.~\ref{fig:lineratio} presents the observed [NeV]/[NeIII] and [OII]/[NeIII] line ratios of \gh~compared to low redshift ($\langle z\rangle=0.137$) \acl{sdss} \citep[\acs{sdss},][]{York:00}/\acl{boss} \citep[\acs{boss},][]{Eisenstein:11}{}{} data. No selection is made in redshift hence the mean redshift is a survey selection effect. The measured line strengths are taken from the catalogue of emission line properties from \cite{Thomas2013}\footnote{\url{www.sdss4.org/dr17/spectro/galaxy_portsmouth}}. We restrict this comparison to galaxies with [OIII] and [NeIII] detected at $\ge 2.5\sigma$ and [NeII] detected at $\ge 2\sigma$. The uncertainties on the flux ratios are smaller than the hexagonal bins. No selection is made for galaxy type and this sample includes both AGN and star-forming galaxies as selected by the \citet[][hereafter BPT]{Baldwin:81} criteria. We applied no reddening correction as all three lines are close in wavelength and hence their ratios are minimally affected by dust absorption. We also examine the parameter space of the emission line ratios from \ac{llagn} modelled by \cite{Groves:06}. Finally, we compare the line ratios to those predicted from the MAPPINGS III library of fast radiative shock models \citep{Allen:08}. We plot the radiative shock and photoionised precursor (sp) models  and the simpler shock (s) or precursor (p) only models for a low, \ac{smc}-like metallicity (comparable to the metallicity estimates from the \acs{beagle} fit \S\ref{sec:res:beagle}) although the general result holds for nearly all metallicity templates.

\gh~lies in a region of parameter space below and to the left of the bulk of the \ac{sdss} sources, clearly due to the strong [NeIII] line, but to the right of the \ac{llagn} parameter space.  \gh~is  consistent with the sp model from MAPPINGS III. This result is driven by the observed [OII]/[NeIII] line ratio. The MAPPINGS III models are a function of velocity, magnetic parameters, preshock density and abundances. The range of velocities modelled only reaches up to 1,000\,km\,s$^{-1}$, but the models find that for the sp models, [NeV]/[NeIII] correlates with velocity and [NeV]/[NeIII] is consistent with the fastest velocities. Given the uncertainties in the observed line ratios and the overlapping nature of the sp models in this line ratio parameter space, it is not possible to place any constraints on the magnetic parameters, preshock density or abundance. While typical EELR galaxies have [NeV]/[NeIII] line ratios $>1$ \citep{Fu:09a}, which is inconsistent with \gh, the suggestion that \gh~is powered by shock emission is consistent with earlier work on compact radio galaxies \citep{Best:00b,DeBreuck:00,Inskip:02b}.

Recently, studies of [NeV]-selected galaxies \citep{Cleri:23a} have shown that $z\approx 2$ galaxies in CANDELS with G102/G141 WPC3 grism spectra have line ratios similar to the local SDSS galaxies in Fig.~\ref{fig:lineratio}. \cite{Cleri:23b} has also shown that galaxies observed with the NIRSpec instrument on board {\it JWST} have line ratios consistent with low-luminosity AGN. Therefore, \gh~has line ratios different from both these samples, further suggesting a unique physical mechanism compared to the general radio-quiet AGN population. 

The presence of the [NeV] line suggests an ionisation potential strong enough to ionise helium into He$^{\rm 2+}$. As \gh~lies at $z\approx 3$, towards the end of the epoch of Helium reionisation, this source, and other compact radio-loud AGN, could be great targets for {\bf observations of potential} absorption by the hyperfine transitions of He$^{\rm 3+}$ (rest-frame frequency of 8.66\,GHz) as well as neutral hydrogen. Future spatially resolved, $0.1''$, spectroscopy with {\it JWST} of the [NeV], [NeIII] and [OII] lines could map out the origin of the photoionisation across and in front of the jet. Observed asymmetry in the Lyman$\alpha$ line could also be used to model line of sight absorption by HI, providing an independent comparison to HI absorption seen in the radio. 

\subsection{Black Hole Mass and Accretion Rate Constraints}

\subsubsection{Bolometric Luminosity}
\label{sec:dis:bol}

Recently, \cite{Barchiesi:24} has investigated [NeV]-selected galaxies at $0.6<z<1.2$ in {\bf the} \acl{cosmos} \citep[\acs{cosmos};][]{scoville:07}. They found a tight correlation between the AGN bolometric luminosity derived from broad-band SED fitting and the luminosity of the [NeV] line. Here we use this relation to estimate the AGN bolometric luminosity of \gh, finding a value of $1.1\pm 0.4\times 10^{12}\,$L$_\odot$ where the uncertainty comes from the uncertainty of the [NeV] flux measurement (i.e. does not include any scatter in the relation between AGN bolometric luminosity and [NeV] line luminosity). From the \acs{wise} non-detections presented in S22 we can derive an upper limit to the rest-frame mid-IR luminosity and hence to the AGN bolometric luminosity, $L_{\rm BOL}$. We use the conversion from $3.4\,\mu$m luminosity to bolometric luminosity given in \cite{kim:23}. From this we find $L_{\rm BOL}\le 1.7\times 10^{12}\,L_\odot$ (reported in Table~\ref{tab:sum_opt}). This value is consistent with the bolometric luminosity derived from the [NeV] line luminosity.

The luminosity of the [NeV] line is also known to correlate with the unobscured X-ray luminosity \citep{mignoli:13}. A typical value for the X-ray to [NeV] luminosity ratio, X/[NeV], is $\approx 400$ for unobscured Seyferts and QSOs \citep{gilli:10} albeit with an order of magnitude scatter. We estimate an unobscured X-ray luminosity of $L_{\rm X}^{\rm [NeV]}\approx 1.2\times 10^{44}\,$erg\,s$^{-1}$. This result is consistent with the X-ray luminosity upper limit inferred from eROSITA ($L_{\rm X}<1.7\times 10^{44}\,$erg\,s$^{-1}$ assuming a photon index of $\Gamma=1.8$) although compact radio sources typically show large absorbing column densities, $N_{\rm H}\approx 10^{21}$-$10^{24}\,$cm$^{-2}$ \citep{Sobolewska:19}.

\subsubsection{Black Hole Mass and Eddington Accretion}
\label{sec:dis:mass}
Assuming that the [NeV] bolometric luminosity is an approximation of the true value, we can determine rough constraints on the black hole mass and accretion rate by using the jet power calculated in \S\ref{sec:res:jet}. We use analytical solutions to the \cite{Blandford:77}, hereafter BZ, model to express the jet power as function of black hole mass, $M_{\rm BH}$, Eddington accretion rate, $\dot{m}$ and the dimensionless black hole spin parameter, $a$ \citep[i.e. we assume a spinning][black hole]{Kerr:63}. This dimensionless spin parameter is limited to $-1<a<1$ with a negative value corresponding to retrograde spin of the accretion disk relative to the black hole {\bf spin}. The spin paradigm of the BZ model, where the magnetic fields thread the black hole, posits that the energy of the jet comes from the angular momentum of the black hole not the accretion. Even if some of the magnetic field threads the accretion disk \citep[i.e. the BP model of][]{Blandford:82}, this magnetic field is still a function of the spin as the rotating general relativistic metric of the Kerr black hole `frame-drags' the magnetic field and inner accretion disk \citep{Meier:99}.  

We follow the formalism for jet power, $Q_{\rm jet}$, developed in \cite{Meier:02} which is used frequently in the literature \citep[e.g.][]{Fanidakis:11,amarantidis:19}. While \cite{Meier:01} has formulas for the jet power which depend on extra factors, such as the disk viscosity and relative angular momentum, very similar jet powers are estimated for reasonable values of these extra (but unmeasurable) parameters. Hence, we caution that the expressions below are approximations. 

As the strength of the jet power is largely determined by the strength of the polaroid magnetic field then it depends on the structure of the accretion disk \citep{Livio:99}. Therefore, the expressions used for the jet powers have either a geometrically thin disk \citep[TD, with efficient accretion,][]{Shakura:73}, or the geometrically thick disk \citep[occuring in an advection-dominated accretion flow, ADAF, with lower efficiency accretion,][]{Rees:82}:

\begin{equation}
        Q_{\rm jet}^{\rm TD}\approx 2.5\times 10^{36}\bigg(\frac{M_{\rm BH}}{\rm 10^9\,M_\odot}\bigg)^{1.1}\,\bigg(\frac{\dot{m}}{0.01}\bigg)^{1.2} \times a^2~~{\rm (W)} 
        \label{eq:jet_td}
\end{equation}
\begin{equation}        
   Q_{\rm jet}^{\rm ADAF}\approx 2\times 10^{38}\bigg(\frac{M_{\rm BH}}{\rm 10^9\,M_\odot}\bigg)\,\bigg(\frac{\dot{m}}{0.01}\bigg) \times a^2~~{\rm (W).} 
        \label{eq:jet_adaf}
\end{equation}

\noindent
Typically accretion disks are believed to transition from an ADAF to a TD as $\dot{m}$ increases through 0.01-0.03 \citep[based of typical values of $\dot{m}$ seen in the transition from ADAFs to TDs in Galactic black holes][]{vm:19}.

We do not have a constraint on $\dot{m}$ or $M_{\rm BH}$ for \gh, but we have a joint constraint on the black hole mass and $\dot{m}$ from the bolometric luminosity (from the definition of the Eddington accretion rate: $\dot{m}=L_{\rm BOL}/(k\times M)$ where $k$ depends on the units used). Using the estimate of the bolometric luminosity {\bf derived in \S\ref{sec:dis:bol}} from the [NeV] line ($\approx 1.1\pm 0.4\times 10^{12}\,$L$_\odot$) we get:

\begin{equation}
    \bigg(\frac{M_{\rm BH}}{\rm 10^9\,M_\odot}\bigg)\,\bigg(\frac{\dot{m}}{0.01}\bigg) \approx 3.2\pm 1.2
\end{equation}

Reconciling this result and the high jet power, $\sim 1.5\times 10^{39}\,$W (from \S\ref{sec:res:jet}), with equations~\ref{eq:jet_td} and~\ref{eq:jet_adaf} implies values of $a^2\gg1$. However, the dimensionless black hole spin parameter is limited to $\le 1$ for the \cite{Kerr:63} solution for black holes, hence our observations are not consistent with equation~\ref{eq:jet_td}. In the ADAF case (equation~\ref{eq:jet_adaf}) it would imply $a^2\approx 2.1$, i.e. only just above unity.  Therefore, allowing for the uncertainties on the jet power and/or the bolometric luminosity, this result could be consistent with the ADAF solution, i.e. realistic values of $a\lesssim 1$ could be achieved, but not the TD solution.  While the [NeV] constrain on the bolometric luminosity has some uncertainty, the upper limit from the {\it WISE} non-detection, $<1.7\times 10^{12}\,$L$_\odot$, is more robust and equally well supports the conclusions above. The conclusion that the accretion disk is in ADAF and not a thin disk state is consistent with the interpretation from \S\ref{sec:res:hst:spec} that \gh~is a LERG and not a HERG. Even using the equations from \cite{Meier:01} we find the only consistent solution is the ADAF one.

The inference that the BH has an ADAF disk at first may seem inconsistent with high-excitation state inferred by the [OII] luminosity (see \S\ref{sec:res:hst:spec}). However, the [OII] luminosity has an unknown contribution from star formation as the SFR derived from SED fitting provides a poor constraint (see \S\ref{sec:res:beagle}). Furthermore, in radio-loud AGN it has been shown that [OII] is predominantly due to some combination of star-formation, EELR and shocks \citep[][]{Maddox:18}, not the broad or narrow line region, hence is not a good indicator that an AGN has a high enough accretion rate to be in TD state.

If the accretion disk is in an ADAF mode with then we can very tentatively constrain the black hole mass assuming a maximum Eddington accretion of $\dot{m}\le 0.03$. With this value we obtain a constraint on the black hole mass of $\ge 10^9\,$M$_\odot$. In this case the black hole mass would be more than an order of magnitude above the local black hole mass/host galaxy relation of \cite{Reines:15} relation for local AGN hosts (which would predict a black hole mass of just $\approx 6\times 10^7\,$M$_\odot$). However, it is consistent with the \cite{Reines:15} relation for local massive ellipticals which are believed to be the local descendants of powerful HzRGs \citep[e.g.,][]{Seymour:07,DeBreuck:10}. If the [NeV] bolometric AGN luminosity is over-estimated, then the inconsistency with both disk models increases. Hence, the very high jet power of \gh~implies that it likely has a high-spin, massive black hole, $\ge 10^9\,$M$_\odot$ with an ADAF accretion disk, and  $\approx 10^{12}\,$L$_\odot$ bolometric luminosity.

\subsection{Future Evolution of \gh}

The compact nature and spectral properties of GLEAM J0917-0012 suggest that it might be in an early stage of radio galaxy evolution, likely classified as a MPS source or a redshifted GPS source. As GPS sources are thought to represent the progenitors of large-scale radio galaxies \citep{odea:98,orienti:16} the proposed evolutionary scenario \cite[e.g][]{fanti:95,An:12} suggests that these radio sources expand into their host galaxies, interacting strongly with the interstellar medium. We cannot determine if \gh~will (a) keep growing from GPS to CSS to extended radio galaxy, (b) shut off and stop expanding, or (c) become "frustrated" by a dense medium \citep{vanBreugal:84,kawakatu:09}. We can say that the observed jet-induced shocks and extended [OII] emission in \gh~are consistent with the picture of ongoing interaction with the host galaxy \citep{Bicknell:18}. 

The high black hole mass, $\ge 10^9\,$M$_\odot$, combined with the low accretion rate, has significant implications for galaxy evolution models. Observed at roughly 2\,Gyr after the Big Bang, this system coincides with the peak of cosmic star formation and AGN activity \citep{madau:14}. The black hole's mass indicates significant early growth \citep[][]{McAlpine:18} which has now slowed and the accretion disk is in a LERG state. However, even with low accretion rate, a high black hole spin is enough to produce powerful radio jets which appear to be impacting the host galaxy via shocks and potentially strong outflows. 

The host galaxy's SFR (and hence sSFR) is poorly constrained, but we can say with confidence that it is below the main sequence for star-forming galaxies at this redshift (see \S\ref{sec:dis:host}). Could the radio jets be quenching the star-formation? Deep narrow surveys are finding many quenched massive galaxies in the early Universe \citep{Carnall:23}. The power of the jet reported here is more than enough to quench star-formation. If the on-time of these powerful jets is short enough, $<100\,$ky, (i.e. a low radio duty cycle where they rarely grow into large radio galaxies) then they could be too rare to be seen in the deep pencil-beam {\it JSWT} surveys.

Hence, this source potentially demonstrates how AGN feedback might regulate growth and star formation in this critical epoch \citep{Harrison:17}. The source's high redshift provides a valuable opportunity to study this evolutionary process in the early Universe.
Future observations to constrain the star formation history (e.g. is the SFR declining?) and to map out the ionisation state in and around the jet will better constrain feedback.

\section{CONCLUSIONS}
\label{sec:con}
This paper confirms the redshift of the enigmatic radio source \gh~which, due to its compact size and faint $K$-band host, was originally suspected to be $z\ge 6$, but is now revealed to be a  young and compact radio galaxy at cosmic noon. The main results of this work are:

\begin{enumerate}
    \item The 1D WFC3 spectrum detects the [OII]$\lambda$3727 doublet as well as the [NeV]$\lambda$3426 and [NeIII]$\lambda$3867 emission lines, confirming the redshift to be $z=3.004\pm 0.001$. The 2D spectrum shows extended [OII] emission ($\approx 4.8\,$kpc) aligned with the two components seen in the high resolution ($60\times 20$\,mas) LBA imaging at 2.276\,GHz. 
    \item The detection of Neon lines suggest the presence of an AGN powered extended emission line region as seen in many other radio galaxies. However, the scale of this region in \gh~is smaller than the typical $\geq 10\,$kpc scales. 
    \item The nearby optical source is found to be at $z=2.13$ and is hence unrelated to \gh~despite its chance alignment with the radio jets. 
    \item SED fitting with {\tt BEAGLE} finds a stellar mass of $\approx 2\times 10^{11}\,$M$_\odot$, making \gh~lie at the low end of the 500\,MHz luminosity, $L_{\rm 500MHz}$, and stellar mass distribution of the most extreme radio galaxies at similar redshifts. It has a similar projected size ($<10\,$kpc) and $L_{\rm 500MHz}$ to just one other classic HzRG: WN\,J1115$+$5016. 
    \item The extended component of the [OII] line in \gh~is reminiscent of the jet-cloud interactions observed in local 3C CSS sources. 
    \item Modelling of the radio emission with {\tt RAiSE} along with constraints on the galaxy size from the F140W image provides measures of the jet power and age of this radio source. Unsurprising, this source is young, $\approx 50\,$kyr with high jet power of $\approx 10^{39}\,$W.
    \item The crude constraint on the bolometric luminosity, $\approx 10^{12}\,$L$_\odot$, combined with the estimate of the radio jet power tentatively implies a highly spinning central black hole in an ADAF state with a relatively low accretion rate ($\lambda_{\rm EDD}\le 0.03$) and a high mass, $\ge 10^{9}\,$M$_\odot$. This result is consistent with LERG classification of this radio galaxy based on its [OII] and radio luminosity.
    \item The [NeV]$\lambda$3426/[NeIII]$\lambda$3867 vs [OII]$\lambda$3727/[NeIII]$\lambda$3867 line ratios are most consistent with radiative shock models with precursor photoionisation. Hence, \gh~is a young radio galaxy with compact jet inducing shock ionisation. 
\end{enumerate}

We summarise that \gh~is a prime example of a young radio galaxy interacting via shocks in the ISM of its host galaxy as demonstrated by the close alignment of the radio jet and [OII] emission as well as modelling of the emission line ratios. Potentially \gh~is on the way to having its star-formation quenched. If the jet is short-lived then this mechanism could account for the numerous quenched massive galaxies seen in the narrow deep surveys. Such a source is a prime target for a plethora of follow-up observations investigating the nature of interaction of the jet with the host galaxy.  Observations of other extreme radio to $K$-band flux ratio sources with low-frequency spectral curvature \citep[e.g.][]{Broderick:24} could provide numerous more candidates for vigorous mechanical feedback at cosmic noon.

\begin{acknowledgement}
The authors acknowledge the Noongar people as the traditional owners and custodians of Whadjuk boodjar, the land on which the majority of this work was completed. NS thanks Brent Groves and Mara Salvato for useful discussions. We thank the reviewer for their very useful and constructive comments on the paper.

The work of DS was carried out at the Jet Propulsion Laboratory, California Institute of Technology, under a contract with NASA.

This research is based on observations made with the NASA/ESA {\it Hubble Space Telescope} obtained from the Space Telescope Science Institute, which is operated by the Association of Universities for Research in Astronomy, Inc., under NASA contract NAS 5–26555. These observations are associated with programs 16184 and 16662. TA acknowledges the grant support through National SKA Program of China (2022SKA0120102)

This work has made use of data from the European Space Agency (ESA) mission {\it Gaia} (\url{https://www.cosmos.esa.int/gaia}), processed by the {\it Gaia} Data Processing and Analysis Consortium (DPAC, \url{https://www.cosmos.esa.int/web/gaia/dpac/consortium}). Funding for the DPAC has been provided by national institutions, in particular the institutions participating in the {\it Gaia} Multilateral Agreement.

The Long Baseline Array is part of the Australia Telescope National Facility (\url{https://ror.org/05qajvd42}) which is funded by the Australian Government for operation as a National Facility managed by CSIRO. This work was supported by resources provided by the Pawsey Supercomputing Centre with funding from the Australian Government and the Government of Western Australia. 

The Hyper Suprime-Cam (HSC) collaboration includes the astronomical communities of Japan and Taiwan, and Princeton University. The HSC instrumentation and software were developed by the National Astronomical Observatory of Japan (NAOJ), the Kavli Institute for the Physics and Mathematics of the Universe (Kavli IPMU), the University of Tokyo, the High Energy Accelerator Research Organization (KEK), the Academia Sinica Institute for Astronomy and Astrophysics in Taiwan (ASIAA), and Princeton University. Funding was contributed by the FIRST program from the Japanese Cabinet Office, the Ministry of Education, Culture, Sports, Science and Technology (MEXT), the Japan Society for the Promotion of Science (JSPS), Japan Science and Technology Agency (JST), the Toray Science Foundation, NAOJ, Kavli IPMU, KEK, ASIAA, and Princeton University. 

This paper is based [in part] on data collected at the Subaru Telescope and retrieved from the HSC data archive system, which is operated by the Subaru Telescope and Astronomy Data Center (ADC) at National Astronomical Observatory of Japan. Data analysis was in part carried out with the cooperation of Center for Computational Astrophysics (CfCA), National Astronomical Observatory of Japan. The Subaru Telescope is honored and grateful for the opportunity of observing the Universe from Maunakea, which has the cultural, historical and natural significance in Hawaii. 

 This research made use of Astropy\footnote{http://www.astropy.org} a community-developed core Python package for Astronomy \citep{astropy:2013, astropy:2018}. This research made use of Photutils, an Astropy package for detection and photometry of astronomical sources \citep{Bradley:24}. This research made use of PetroFit \citep{Geda:22}, a package based on Photutils, for calculating Petrosian properties and fitting galaxy light profiles.

 This research made use of ds9, a tool for data visualization supported by the Chandra X-ray Science Center (CXC) and the High Energy Astrophysics Science Archive Center (HEASARC) with support from the JWST Mission office at the Space Telescope Science Institute for 3D visualization. 
\end{acknowledgement}


\bibliography{bigbiblio}


\end{document}